\documentclass[%
    preprint,
 amsmath,amssymb,
pra,
]{revtex4-2}

\usepackage[utf8]{inputenc}
\usepackage[utf8]{inputenc}%
\usepackage[dvipsnames]{xcolor}
\usepackage{graphicx}%
\usepackage[normalem]{ulem}
\usepackage{todonotes}

\let\vec\boldsymbol

\newcommand{\todoNL}[2][]
{}

\usepackage{hyperref}%
\hypersetup{
    colorlinks=true,
    linkcolor=blue,
    filecolor=magenta,      
    urlcolor=cyan,
    citecolor=cyan,
    pdfpagemode=FullScreen,
    }

\begin{document}

\title{Extended locally monochromatic approximations of Strong-Field QED processes}

\newcommand{\HIJ}{Helmholtz Institute Jena, Fröbelstieg 3, 07743 Jena, Germany}
\newcommand{\GSI}{GSI Helmholtzzentrum für Schwerionenforschung GmbH, Planckstraße 1, 64291 Darmstadt, Germany}
\newcommand{\IOQ}{Institute of Optics and Quantum Electronics, Friedrich-Schiller-Universität, Max-Wien-Platz 1, 07743 Jena, Germany}

\author{Nikita Larin}
\email{nikita.larin@uni-jena.de}
\affiliation{\HIJ}
\affiliation{\GSI}
\affiliation{\IOQ}

\author{Daniel Seipt}
\affiliation{\HIJ}
\affiliation{\GSI}
\affiliation{\IOQ}

\date{\today}

\begin{abstract}
Strong-field QED (SFQED) probability rates in the locally monochromatic approximation (LMA) have become an indispensable tool for simulations of processes like gamma-ray emission or electron-positron pair production in laser-particle collisions. We revisit the LMA derivation and explicitly demonstrate that it is based on the separation of time scales, neglection of the long-range interference effects and subsequent averaging over the cycle scale. Doing so, we obtain unambiguously LMA rates for arbitrary polarizations of the plane wave background. Additionally, we partially restore the finite bandwidth effects that are lost in the LMA derivation. The bandwidth-restored result we refer to as the LMA$^+$ and show that it agrees with the full SFQED predictions better than the standard LMA. We use LMA$^+$ to address previously inaccessible observables and formulate a new limitation on the applicability of locally monochromatic approximations in general. We provide analytical results for the angular-integrated LMA$^+$ probability rate and the fully differential probability that account for the finite bandwidth effects. 

\end{abstract}

\maketitle

    \section{Introduction}

    The dynamics of charged particles in ultra-strong electromagnetic fields is characterized by nonlinear effects \cite{DiPiazza:RevModPhys2012, Gonoskov2022, Fedotov2023}. These nonlinearities can, for instance, substantially alter the characteristic signatures of the emitted radiation \cite{Bula1996,  Chen1998, Yan2017, Mirzaie2024}, facilitate radiation back reaction \cite{Cole2018, Poder:PRX2018, Blackburn:RevModPlasma2020,sikorski_novel_2024}, or lead to the generation of copious amounts of matter-antimatter pairs \cite{Bell:PRL2008a}. Such highly nonlinear regimes have now become experimentally accessible by employing ultra-high intensity laser pulses \cite{Danson2019, Yoon2021}. In general, SFQED effects are important for plasmas in extreme-field conditions not only in laboratories but also in astrophysical environments \cite{Zhang2020, Gonoskov2022, Slade-Lowther:NJP2019, cruz_coherent_2021}. Dedicated experiments to study SFQED processes in detail can be performed by colliding high-intensity laser pulses with ultrarelativistic electron bunches, either from a laser plasma accelerator \cite{Cole2018, Poder:PRX2018, Mirzaie2024} or from a conventional accelerator  \cite{Bamber:PRD1999, Abramowicz2021, Abramowicz2024, ChenE320_2022}. 
    
    The strength of laser matter interaction is usually characterized using two parameters: (i) the classical nonlinearity parameter $a_0$ and (ii) the quantum parameter $\chi$. The parameter $a_0 = |e|E/m\omega c$ represents the work done by an electric field with field strength $E$ on a particle with charge $e<0$ over a distance set by the wavelength of the field $\lambda = 2\pi c/\omega$ in units of the particle's rest energy $mc^2$. Alternatively, if one implies a quantum-mechanical picture, $a_0$ tells how many field ``photons'' participated in a SFQED process. So, if it becomes of the order of one or larger, we expect to observe features of a nonlinear interaction between a charged particle and a laser field \cite{Fedotov2023, Gonoskov2022}. It also serves as the inverse Keldysh parameter \cite{Popruzhenko_2014}, meaning that the processes becomes perturbative as $a_0\to0$ and turns quasi-static as $a_0\to \infty$. The quantum parameter is defined as $\chi =e\hbar\sqrt{-\left({F}_{\mu\nu}p^\nu\right)^2}/m^3c^4$, where ${F}_{\mu\nu}$ is the electromagnetic field strength tensor and $p^\mu$ is the particle's four-momentum. Quantum processes become efficient if $\chi\sim 1$ \cite{Ritus1985}. For an electron emitting a photon, this means that the latter acquires a significant fraction of the electron's initial energy. The photon itself is characterized by its own $\chi_\gamma$, and it is efficiently converted into $e^+e^-$-pairs if $\chi_\gamma\sim 1$.

    When $\chi$ {exceeds} the order of unity one usually has to rely on computer simulations to describe the outcome of experiments, especially if emitted photon's $\chi_\gamma$ exceeds unity, and they efficiently produce a second generation of charged particles. These, in turn, keep radiating and produce subsequent generations, forming so-called shower-type cascade \cite{Sokolov2010, Bulanov2013}. For each subsequent generation the total energy is shared by more and more particles, and eventually the $\chi$ parameter becomes significantly less than one, and the cascade stops \cite{Blackburn2019,pouyez_multiplicity_2024,pouyez_kinetic_2025}, unless there is an efficient re-acceleration of the particles in which case the cascade growth is self-sustained \cite{Bell:PRL2008a, Fedotov:PRL2010, Bulanov:PRL2010, mercuri-baron_growth_2025, Seipt2021}.

In particular, shower-type cascades with medium-to-high multiplicity of final state particles are of the great importance for contemporary experimental campaigns of SFQED \cite{Abramowicz2021,Abramowicz2024,seipt_nonlinear_2025}. Unfortunately, a complete analytical treatment of these processes is near impossible within the full SFQED framework due to the complexities of high-order S-matrix calculations \cite{Fedotov2023}. However, for sufficiently strong and/or long pulses, it is possible to split the higher order processes %
into first order building blocks \cite{Dinu2019,Dinu2020,Torgrimsson2021}, such as nonlinear Compton scattering (NCS) \cite{nikishov1964quantum} and nonlinear Breit-Wheeler (NBW) pair production \cite{ReissNBW1962}. In doing so, one neglects all contributions from off-shell intermediate particles \cite{Fedotov2023}. But it opens avenues for numerical simulation frameworks in which the particles propagate on classical trajectories between the quantum processes, described by some probability rates \cite{Gonoskov2022}.

The rates for the quantum processes in arbitrary field configurations, such as the ones found in laser-plasma interactions, are the most commonly employed in the locally constant field approximation (LCFA) \cite{Ritus1985}, or its extensions \cite{DiPiazza2018, DiPiazza2019, Ilderton2019, King2020, Gelfer2022, Lv2021}. The applicability of the LCFA is, however, limited to $a_0\gg1$ and $a_0^3/\chi\gg1$   \cite{ DinuPRL2016, Blackburn2018, DiPiazza2018}. {Many features like harmonic and sub-harmonic structure of the spectra are lost in the LCFA. Moreover, for the NCS the LCFA overestimates the low-energy part of the emitted spectrum, even when all forementioned requirements are fulfilled.} Despite these shortcomings, the LCFA is nowadays implemented in many particle-in-cell simulation codes' SFQED modules \cite{Ridgers2014, Gonoskov2015, Vranic2016, Lobet2016, Derouillat2018}. 
    
For collisions of a high-intensity laser with a particle beam, where the laser field can be assigned a distinct wave vector and pulse envelope, such that the pulse contains many carrier wave cycles, another approach is the \emph{locally monochromatic approximation} (LMA). In contrast to the LCFA, the LMA is not restricted to large $a_0$, and is able to resolve harmonics in the emitted spectrum. Incidentally, the LMA was actually used for the simulations of the SLAC E-144 experiment \cite{Bamber:PRD1999}, by just using the results for the SFQED scattering rates in infinite monochromatic plane waves modified by a local value of $a_0$. The LMA was formally derived only relatively recently by performing a separation of carrier wave and envelope time scales \cite{Heinzl2020}. Nevertheless, the LMA has already proven its utility and reliability for numerical simulations, especially in the transition regime $a_0\sim 1$ \cite{Blackburn2021, Tang2022, Nielsen2022, Blackburn2022, Blackburn2023, Tang2023}.

In this paper, we revisit the derivation of LMA and find a novel procedure to obtain the LMA probability rates from the full SFQED probability. The essential aspect in our approach is that a cycle-averaging procedure is necessary for finding the correct LMA rate. While this seems like a natural necessity in view of the scale separation, and it was implied in the application of the LMA in simulation codes \cite{Blackburn2021}, it was never explicitly employed in the derivation of the LMA rates themselves. By manifestly performing the cycle-averaging, we are able to fill several gaps in the LMA derivation that previously required numerical arguments. In addition, we propose an {extension} of the LMA, which we call LMA$^+$, that restores some bandwidth effects and removes divergencies in the fully differential LMA rate and probability. We confront our findings with the standard LMA, the LCFA, and with exact SFQED $S$-matrix calculations to demonstrate the validity of our new results and discuss the underlying physics. Our findings indicate an additional limitation for the applicability of the LMA for large $a_0\gg1$. Moreover, we find new analytical results for both the angularly integrated LMA$^+$ rates, and the LMA$^+$ probabilities when integrated over the complete laser pulse history.

    The paper is organized as follows. In Sec.~\ref{sect:lma} we derive the LMA probability rates for arbitrary polarizations of a plane wave background, by explicitly employing the cycle-averaging procedure. In Sec.~\ref{sec:LMA_plus} we formulate fully differential LMA$^+$ probability rate and show how to obtain analytical expressions for the angular-integrated LMA$^+$ and fully differential probability that account for the finite bandwidth effects. The LMA$^+$ results are then compared with the standard LMA, the LCFA and the full SFQED calculations (Sec.~\ref{sect:lma+numeric}), and further analytic results are presented (Sec.~\ref{sect:lma+analytic}). We conclude in Sec.~\ref{sec:summary}. In Appendix~\ref{sec:app.A} we provide a detailed derivation of the series representation of the floating average, which we use in our LMA derivation. Finally, in Appendix~\ref{sec:app.B} we show evaluation of the generalized Neumann-type integrals that appear in the cycle-averaging procedure. Throughout this article, we employ Heaviside-Lorentz natural units with $\hbar=c=\epsilon_0=1$, such that the fine structure constant reads $\alpha=e^2/4\pi$. Scalar products between four-vectors are denoted as $k\cdot p=k^\mu p_\mu$.

\section{New Derivation of the LMA} \label{sect:lma}

    The locally monochromatic approximation (LMA) has become a cornerstone of numerical simulations of SFQED processes in collisions of particle beams with high-intensity lasers, especially in the transition regime $a_0\sim1$, where the LCFA is not applicable. But, it is more numerically expensive and applicable only for the backgrounds with characteristic carrier frequency scale (e.g., laser frequency $\omega$) and which are sufficiently long. For a more detailed discussion of the requirements we refer to Ref.~\cite{Heinzl2020}, where the first formal derivation of the LMA was presented by approximating on the level of the strong-field S-matrix.
    As we will see further, the requirement for the existing of two different time scales is crucial in the LMA derivation, since it allows performing the cycle-averaging over the fast component of the field. Even though the cycle-averaging was implied and implemented in numerical simulations, it has not yet been employed explicitly in the LMA derivation. Here we will fill this gap in a novel derivation of the LMA, which also opens the avenue for extensions of the LMA.

\subsection{Probability for Nonlinear Compton Scattering} \label{subs:nlc}

\begin{figure}[h!]
    \centering
   \includegraphics[width=0.5\linewidth]{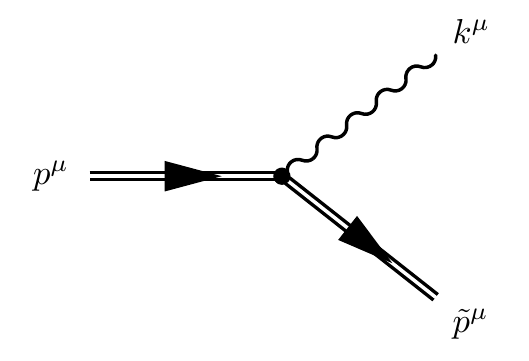}
    \caption{SFQED Feynman diagram for Nonlinear Compton Scattering.}
    \label{fig:diagram}
\end{figure}

To demonstrate the new derivation of the LMA, we specifically consider nonlinear Compton scattering (NCS). An initial electron with four-momentum $p^\mu$ collides with a plane electromagnetic wave with the normalized vector potential $\vec{a}_\bot\left(\kappa\cdot x\right) = a_0\vec{f}_\bot\left(\kappa\cdot x\right)$ and four-wavevector $\kappa^\mu$. The symbol "$\bot$" stands for the components perpendicular to the background plane wave propagation direction. In the course of interaction, the electron emits a photon with the four-momentum $k^\mu$ and propagates further with the final four-momentum $\tilde{p}^\mu$ (see, Fig.~\ref{fig:diagram}).

The differential probability for the process, averaged over the initial electron's spin and summed over the all final spin and polarization states of the final electron and photon, may be written as \cite{Dinu2013}
\begin{align}
    \frac{d\mathbb{P}}{d\ell d^2\vec{\rho}_\bot} &=- \frac{\alpha}{\pi^2}{A}\iint\limits_{\mathbb{R}^2}\left[1 +  a_0^2{B}\theta^2 \langle\vec{f}'_\bot\rangle^2\right]\exp\left[{i\ell\theta{G}\left(\varphi,\theta\right)}\right]d\varphi d\theta
    \label{Eq.2.1},
    \end{align}
where we introduced the shorthand notations
\begin{align}
A  &= \frac{\ell }{{\left(1 + \vec{\rho} _ \bot ^2 + 2\eta\ell \right)^2}}\,,\\
B &=\frac{1}{2}+\frac{\eta^2\ell^2}{\left(1 + \vec{\rho}^2_\bot\right)\left(1 + \vec{\rho}_\bot^2 + 2\eta\ell\right)} \,,
    \label{Eq.2.2} \\
{G}\left(\varphi,\theta\right) & = 1 + \frac{a^2_0\langle\vec{f}^2_\bot\rangle + 2a_0\vec{\rho}_\bot\langle\vec{f}_\bot\rangle}{1 + \vec{\rho}^2_\bot} \,.
    \label{Eq.2.3}
\end{align}
Here, we parametrize the final state photon {via the normalized momentum $\ell = k\cdot p/\kappa\cdot \tilde{p}$ that needs to be absorbed from the background field in order to put all particles on-shell, such that $p^\mu + \ell \kappa^\mu = k^\mu + \tilde p^\mu$. Furthermore, we introduce the scaled transverse momentum} $\vec{\rho}_\bot =\left(\vec{k}_\bot -s\vec{p}_\bot\right)/ms$, where $s = \kappa\cdot k/\kappa\cdot p$ is a light-front momentum fraction. {These dimensionless variables are related as $s = 2\eta\ell/(1+2\eta\ell +\vec\rho_\bot^2) $.} The background pulse profile function $\vec{f}_\bot\left(\varphi\right) = g\left(\varphi/\Delta\right)\vec{h}_\bot\left(\varphi\right)$ is a product of an envelope $g\left(\varphi/\Delta\right)$ and a carrier $\vec{h}_\bot\left(\varphi\right)$, where $\Delta$ is a pulse duration. The envelope rapidly vanishes at infinity $g\left(\pm\infty\right)\to0$ and satisfies the condition $g\left(0\right) = 1$.

{The probability, Eq. \eqref{Eq.2.1}, is expressed as a two-fold integral over the average laser phase $\varphi=(\phi+\phi')/2$ and the interference window $\theta=\phi'-\phi$ \cite{Fedotov2023}, where $\phi$ and $\phi'$ are the phase variables of the strong-field $S$-matrix and its complex conjugate, respectively. These variables enter the probability via the floating average of the background profile function
\begin{align}
    \langle \vec{f}_\bot\rangle=\frac{1}{\theta}\int\limits_{\varphi-\theta/2}^{\varphi+\theta/2}\vec{f}_\bot\left(\tilde{\varphi}\right)d\tilde{\varphi} \,,
    \label{Eq.2.4}
\end{align}
where their nontrivial dependencies make an exact analytical treatment intractable if one accounts for the finite duration of the background field.} Usually, either some numerical approaches \cite{Boca2009,Seipt2011, Krajewska2012} or various approximations schemes \cite{Mackenroth2011,Seipt2013,Seipt2016} have to be employed to evaluate the NCS probability, Eq.~\eqref{Eq.2.1}, further.

\subsection{Definition of a Proto-Rate}

{In this subsection we make the first steps towards the LMA probability rate, by first deriving a \emph{proto-rate}, $\mathcal{R}\left(\varphi\right)$, as the quantity that returns the probability when integrated over the laser phase $\varphi$
\begin{align}
\frac{{d}\mathbb{P}}{{d}\ell{d}^2\vec{\rho}_\bot} = \int\limits_{-\infty}^{+\infty} \! d\varphi \:\mathcal{R}\left(\varphi\right) \,.
    \label{eq:protorate}
\end{align} 
In general, the quantity $\mathcal{R}\left(\varphi\right)$ is not strictly positive and therefore \emph{cannot} be interpreted as a probability rate directly. We will demonstrate further below that only under certain {(slowly varying envelope and local) approximations} and, most importantly, a subsequent cycle-averaging procedure, the proto-rate $\mathcal R$ can be converted into a proper positive-definite probability rate.}

The central object for our further analysis is the floating average, Eq.~\eqref{Eq.2.4}, since it contains all information regarding the structure of the plane wave background. To better see how this information is encoded, we rewrite the integral {Eq.~\eqref{Eq.2.4} using the following} series representation (for detailed derivation, see Appendix~\ref{sec:app.A}): 

\begin{align}
\langle \vec{f}_\bot\rangle = \sum_{n=0}^\infty\frac{1}{\left(2n+1\right)!}\left(\frac{\theta}{2}\right)^{2n}\frac{d^{2n}\vec{f}_\bot\left(\varphi\right)}{d\varphi^{2n}}.
    \label{Eq.2.5}
\end{align}
The main advantage of this representation is that it disentangles the dependency on the variables $\varphi$ and $\theta$ from the integral limits.
 
 Given that our background profile function is a product of envelope and carrier functions $\vec{f}_\bot\left(\varphi\right)=g\left(\varphi/\Delta\right)\vec{h}_\bot\left(\varphi\right)$, we use the Leibniz rule to write the $2n^{\mathrm{th}}$-order derivative as follows

\begin{align}
\vec{f}_\bot^{\left(2n\right)}\left(\varphi\right) = \sum_{k=0}^{2n}\binom{2n}{k}g^{\left(k\right)}\left(\frac{\varphi}{\Delta}\right)\vec{h}_\bot^{\left(2n-k\right)}\left(\varphi\right).
    \label{Eq.2.6}
\end{align}
To demonstrate the interplay between the envelope and the interference window $\theta$, we explicitly write down the result for the first two terms $k=0,1$ in the sum (\ref{Eq.2.6}). Substituting (\ref{Eq.2.6}) into (\ref{Eq.2.5}) and 
{exploiting the fact that {components of} the carrier $\vec{h}_\bot\left(\varphi\right)$ are sines or cosines,}
we obtain
\begin{align}
\theta\langle\vec{f}_\bot\rangle\approx 2g\left(\frac{\varphi}{\Delta}\right)\vec{h}_\bot\left(\varphi\right)\sin\left(\frac{\theta}{2}\right) + \frac{1}{\Delta}g'\left(\frac{\varphi}{\Delta}\right)\vec{h}^{\left(-1\right)}_\bot\left(\varphi\right)\left[\theta\cos\frac{\theta}{2} - 2\sin\frac{\theta}{2}\right].
    \label{Eq.2.8}
\end{align}
Here we show the result for the $\theta\langle\vec{f}_\bot\rangle$, since this quantity enters the probability (\ref{Eq.2.1}). The notation $\vec{h}_\bot^{\left(-1\right)}\left(\varphi\right)$ stands for the antiderivative of the carrier and the prime $g'$ corresponds to the derivative with respect to the whole argument. 

 The second term in the expression \eqref{Eq.2.8} contains contributions with two different scalings: $g'/\Delta$ and $g'\theta/\Delta$ (we omit the argument of the envelope's derivative for conciseness). The term $\sim g'/\Delta$ is responsible for the local effects, associated with the gradients of finite envelope, and can be discarded if one employs the slowly varying envelope approximation (SVEA) \cite{Narozhnyi1996}. However, the SVEA alone is insufficient for dropping out the term $\sim\theta g'/\Delta$, which encodes the global structure of the pulse via coupling between the interference window $\theta$ and envelope's gradient $g'$. It corresponds to the interference effects on the scale of the entire pulse and accounts for the fact that the probe particle enters and leaves the interaction region at finite times \cite{King2021, Tang2021}. Therefore, to discard it, we have to neglect the interference on the envelope scale by imposing a local approximation $\theta/\Delta\ll1$. This restriction ensures the inclusion of the interference effects only on the scale of one or several cycles, where the background field deviates insignificantly from the monochromatic plane wave. Thus, given the SVEA and the local approximation ($\theta/\Delta\ll1$), we may rewrite floating averages of the background profile function and its derivative in (\ref{Eq.2.1}) as
\begin{align}
\theta\langle\vec{f}_\bot\rangle\approx 2g\left(\frac{\varphi}{\Delta}\right)\vec{h}_\bot\left(\varphi\right)\sin\left(\frac{\theta}{2}\right),\quad\theta\langle\vec{f}'_\bot\rangle\approx 2g\left(\frac{\varphi}{\Delta}\right)\vec{h}'_\bot\left(\varphi\right)\sin\left(\frac{\theta}{2}\right).
    \label{Eq.2.9}
\end{align}

To obtain the result for $\langle\vec{f}_\bot^2\rangle$ in (\ref{Eq.2.3}) we have to specify the polarization of the carrier explicitly. Henceforth, $\vec{h}_{\bot}\left(\varphi
\right)=\left(\cos\varphi,\delta\sin\varphi\right)$, where $\delta$ is an ellipticity parameter ($\delta = 0$ - linear polarization (LP), $\delta = \pm1$ - circular polarization (CP)). Using this definition, we find
\begin{align}
  \theta\langle\vec{f}^2_\bot\rangle \approx \frac{1}{2}g^2\left(\frac{\varphi}{\Delta}\right)\left[\left({1 + \delta^2}\right)\theta + \left({1 - \delta^2}\right)\cos 2\varphi\sin\theta\right].  
  \label{Eq.2.10}
\end{align}
From (\ref{Eq.2.10}) we see that for {circular polarization} the oscillating term vanishes, leaving only linear dependency on interference window $\theta$.

{To derive our result for the LMA proto-rate, $\mathcal{R}_\mathrm{LMA}$, we substitute all approximations for the floating averages into Eqs.~\eqref{Eq.2.1} and \eqref{Eq.2.3} and apply the generalized Jacobi-Anger expansion \cite{Dattoli1990} with respect to the interference window $\theta$.
We find}
\begin{align}
    \frac{d\mathcal{R}_\mathrm{LMA}\left(\varphi\right)}{d\ell d^2\vec{\rho}_\bot} = -\frac{\alpha}{\pi^2} {A}\sum_{n=-\infty}^{+\infty}\mathcal{D}_n\left(\varphi\right)\int\limits_{-\infty}^{+\infty}\exp\left[i\theta\left(\zeta\left(\varphi\right)-\frac{n}{2}\right)\right] d\theta \,,
    \label{Eq.2.11}
\end{align}
with
\begin{align}
    \zeta\left(\varphi\right) = \ell + \frac{\ell \left(1 + \delta^2\right)}{2\left(1 + \vec{\rho}^2_\bot\right)}\, a_0^2 g^2\left(\frac{\varphi}{\Delta}\right) 
    \label{Eq.2.12}
\end{align}
and
\begin{align}
   \mathcal{D}_n\left(\varphi\right) =  \mathcal{J}_n\left(X,Y\right) + \frac{a_0^2}{2}g^2\left(\frac{\varphi}{\Delta}\right){B} \left[1 + \delta^2-\left(1-\delta^2\right)\cos2\varphi\right]\left(2\mathcal{J}_{n}-\mathcal{J}_{n-2} - \mathcal{J}_{n+2} \right).
    \label{Eq.2.13}
\end{align}
The functions $\mathcal{J}_n$ are generalized {two-argument} Bessel functions \cite{Korsch2006}, which are related to the ordinary Bessel functions {of the first kind} via the series representation \cite{Dattoli1990}
\begin{align}
\mathcal{J}_n\left(X,Y\right) = \sum_{k=-\infty}^{+\infty}J_{n-2k}\left(X\right)J_k\left(Y\right).
    \label{Eq.2.14}
\end{align}
The arguments of the generalized Bessel functions are $X= 2x\cos\left(\varphi-\vartheta\right)$ and $Y = 2y\cos2\varphi$. Here we introduced an azimuthal angle in the transverse momentum plane $\vartheta = \arctan\left(\delta\rho_y/\rho_x\right)$ and shorthand notations
\begin{align}
x = \frac{2\ell \sqrt{\rho_x^2 + \delta^2\rho^2_y}}{1+\vec{\rho}_\bot^2}\, a_0 g\left(\frac{\varphi}{\Delta}\right)\,,\quad 
y = \frac{\ell \left(1 - \delta^2\right)}{4\left(1 + \vec{\rho}_\bot^2\right)}\, a_0^2 g^2\left(\frac{\varphi}{\Delta}\right). 
    \label{Eq.2.15}
\end{align}
For {circular polarization, $\delta=\pm1$,} the generalized Bessel functions \eqref{Eq.2.14} reduce to the ordinary Bessel functions {since $y=0$ in that case.} The remaining $\theta$-integral in Eq.~\eqref{Eq.2.11} can be performed, resulting in the delta distributions:
\begin{align}
    \frac{d\mathcal{{R_\mathrm{LMA}}}\left(\varphi\right)}{d\ell d^2\vec{\rho}_\bot} = -\frac{2\alpha}{\pi} {A}\sum_{n=1}^{+\infty}\mathcal{D}_n\left(\varphi\right)\delta\left[\zeta\left(\varphi\right) - \frac{n}{2}\right] \,.
    \label{Eq.2.16}
\end{align}
{We note that contributions with $n\leq0$ drop from the summation automatically, because of the presence of the delta-distribution and the positiveness of $\zeta(\varphi)$, see Eq.~\eqref{Eq.2.12}.}

\begin{figure}[ht]
    \centering
    \includegraphics[width=0.5\linewidth]{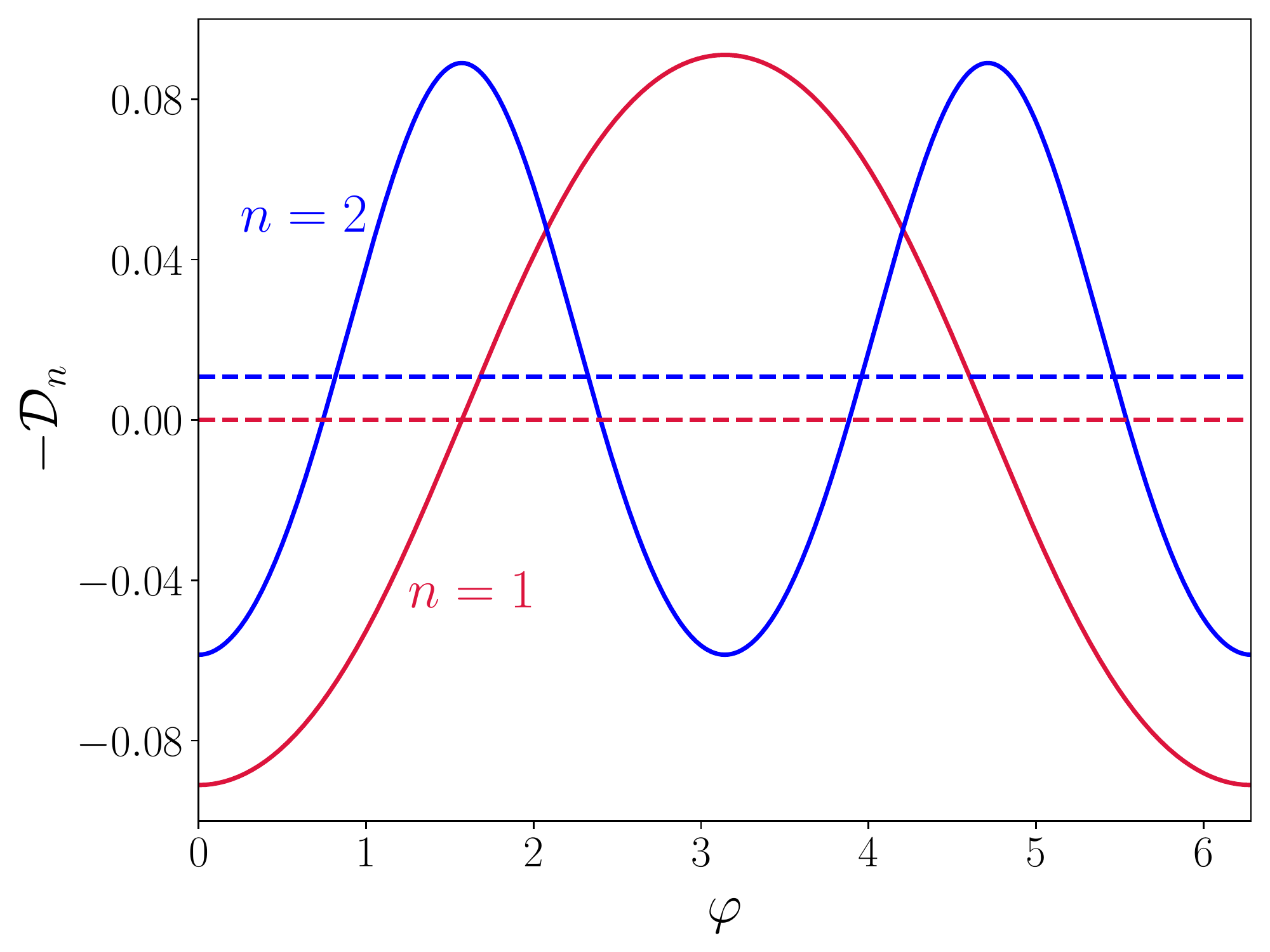}
    \caption{The first two terms of LMA proto-rate ($n=1$ solid red and $n=2$ solid blue lines) in IPW limit. The dashed lines stand for the corresponding cycle-averaged values. The other parameters are chosen as: $a_0=1$, $\eta=0.1$, $\vec{\rho}_\bot=\left(a_0,0\right)$, $\ell = 5/12$ for $n=1$ and $\ell=5/6$ for $n=2$.}
    \label{fig:Proto_Rate}
\end{figure}

{Naively, one might expect that Eq.~\eqref{Eq.2.16} already is the sought after LMA probability rate. But it is not. It still alters in sign, as indicted already at the beginning of the section. Moreover, Eq.~\eqref{Eq.2.16} contains contributions from the \emph{half-integer} harmonics stemming from the condition $\zeta\left(\varphi\right) = n/2$ for odd $n$. To exhibit this behavior explicitly we plot in Fig.~\ref{fig:Proto_Rate} the quantity $- \mathcal D_n(\varphi)$ for $n=0,1$, where we also employ $g\equiv1$, i.e.~the infinite plane wave (IPW) limit. We see that both the $n=1$ (red solid curve) and the $n=2$ (blue solid curve) contributions become negative for some values of $\varphi$. The half-integer harmonic, which corresponds to $n=1$, oscillates around zero and vanishes after averaging over the cycle. Meanwhile, the integer harmonic, $n=2$, averages to some finite positive value which eventually corresponds to the first harmonic in the IPW probability rate \cite{Ritus1985}. In the same manner, all contributions from odd $n$ average to zero, whereas the terms with even $n$ provide the correct result for the IPW rate. These examples show the relevance of the cycle-averaging procedure, which we will demonstrate explicitly for the LMA rate in the next subsection.}

\subsection{Cycle-averaging of the proto-rate}

{In this section, we formally obtain the LMA probability rate unambiguously for circular and linear polarizations of the plane-wave background, by cycle-averaging the LMA proto-rate, Eq.~\eqref{Eq.2.16}, according to
 \begin{align}
 \mathbb{R}_\mathrm{LMA}\left(\varphi\right) \equiv \frac{1}{2\pi}\int\limits_{\varphi - \pi}^{\varphi + \pi}\mathcal{R}_\mathrm{LMA}\left(\varphi'\right)d\varphi' \,.
     \label{Eq.2.24}
 \end{align}
In this procedure we may treat the pulse envelope as a constant, $g\left(\varphi'/\Delta\right)\approx g\left(\varphi/\Delta\right)$, due to the SVEA.
}

{After cycle-averaging (\ref{Eq.2.24}), the proto-rate (\ref{Eq.2.16}) turns into the positive-definite probability rate:
 \begin{align}
 \frac{d\mathbb{R}_\mathrm{LMA}}{d\ell d^2\vec{\rho}_\bot}=-\frac{2\alpha}{\pi} {A} \sum_{n=1}^{+\infty} \mathbb{D}_n\left(\varphi\right)\delta\left[\zeta\left(\varphi\right) - n\right] \,,
     \label{Eq.2.25}
 \end{align}
with 
  \begin{align}   
    \mathbb{D}^{\left(\mathrm{\scriptsize{CP}}\right)}_n\left(\varphi\right) = J_n^2\left(x\right)  +  a_0^2g^2\left(\frac{\varphi}{\Delta}\right){B}\left[2J_n^2 - J_{n+1}^2 - J_{n-1}^2\right]
    \label{Eq.2.26}
\end{align}
for circular polarization and
\begin{align}
    &\mathbb{D}^{\left(\mathrm{\scriptsize{LP}}\right)}_n\left(\varphi\right) = \mathcal{J}_n^2\left(x,y\right)   +\frac{a_0^2}{2}g^2\left(\frac{\varphi}{\Delta}\right){B}\left(2\mathcal{J}_n^2 + \mathcal{J}_{n-2}\mathcal{J}_n + \mathcal{J}_{n}\mathcal{J}_{n+2}-\mathcal{J}^2_{n-1} - 2\mathcal{J}_{n-1}\mathcal{J}_{n+1}-\mathcal{J}_{n+1}^2\right)
    \label{Eq.2.27}
\end{align}
for linear polarization. The integrals that appear in the cycle-averaging are known as Neumann-type integrals, and details on their evaluation are presented in Appendix~\ref{sec:app.B}. Here we just stress that contributions from the half-integer harmonics, which plagued Eq.~\eqref{eq:protorate}, vanish after the cycle-averaging. From now on, the correct numbering of the harmonics is restored, i.e.~$n=1$ refers to the first harmonic etc.}

    The expressions in Eqns.~\eqref{Eq.2.25}-\eqref{Eq.2.27} agree with the known LMA probability rates for circular and linear polarization from the literature, see e.g. Ref.~\cite{Heinzl2020}. The textbook result of the IPW limit can be easily re-obtained from these expressions by just setting $g\equiv 1$. In contrast to the first derivation of the LMA in Ref.~\cite{Heinzl2020}, here we arrived at the same result by approximating the expression for the probability of the process and not the $S$-matrix elements. In contrast to Ref.~\cite{Heinzl2020}, {our LP case result is \emph{not} expressed via a double sum over harmonics that did emerge after squaring the $S$-matrix element.} Numerical studies showed that contribution of the off-diagonal terms $n\neq n'$ in the double sum insignificant for the relevant parameter regime \cite{Heinzl2020}, and some authors used this circumstance for their findings \cite{Nielsen2022, Blackburn2023, Tang2023, King2024_QEDcoupling}. However, it was not clear why there should be such a fundamental distinction between the CP and LP cases.

    {In our new derivation of the LMA rates there are no double sums at all since we take the probability as starting point. However, the cycle-averaging procedure \eqref{Eq.2.24} is absolutely necessary to obtain the LMA probability rates. Incidentally, the integration over the azimuthal angle $\vartheta$ in the transverse momentum plane in the first derivation of the LMA rate in Ref.~\cite{Heinzl2020} for the CP case technically corresponds to the cycle averaging, due to the symmetry of the CP background and the fact that the relevant integrals depend on the laser phase only in the combination $\varphi-\vartheta$. But this symmetry argument only applies for circular polarization. Additionally, we would like to point out that if one performs the cycle-averaging in Eq.~(A34) of Ref.~\cite{Heinzl2020}, the off-diagonal terms are eliminated, and the result coincides with our Eqns.~\eqref{Eq.2.25} and \eqref{Eq.2.27}. Thus, our findings so far are in complete agreement with Ref.~\cite{Heinzl2020}.}
 
    To summarize this section, we emphasize once again the assumptions that we made for deriving the LMA. First, it is necessary to neglect the local gradients of the pulse envelope, employing the SVEA. Second, we discarded the interference effects on the entire pulse scales (local approximation $\theta/\Delta\ll1$). Third, employing once again the SVEA, we performed the cycle-averaging procedure, which turns the sign-alternating proto-rate Eq.~\eqref{Eq.2.16} into the positive-definite LMA probability rate Eq.~\eqref{Eq.2.25}.
    {The implementation of the LMA rates in numerical codes for the simulation of laser-particle collisions is based on a
    splitting the dynamics into slow and fast time scales, where the particles move on cycle-averaged ponderomotive trajectories and all information on the fast quiver oscillations is contained in the LMA probability rates \cite{Blackburn2021, Blackburn2022}. Conceptually, the probability rates entering the simulation should also be the ones taken at the cycle-averaged particle location. Here, we made explicit use of the cycle-averaging procedure in the derivation of LMA rates, rendering the simulation framework more self-consistent.}

    \section{Bandwidth-restored LMA}\label{sec:LMA_plus}

    In deriving the LMA probability rate, we employed a local approximation that effectively discards long-range interference effects associated with the finite extent of the plane wave background \cite{King2021,Tang2021}. This crucial step is what makes the approximation \emph{monochromatic} by removing finite bandwidth effects usually associated to a finite pulse duration. This feature formally manifests in the appearance of delta-distributions in Eq.~\eqref{Eq.2.25}. As a consequence, the LMA probability, which is obtained by integrating Eq.~\eqref{Eq.2.25} over the laser phase $\varphi$ has divergences \cite{Fedotov2023}. Specifically, the triple differential probabilities of the $n$-th harmonic diverge $\propto 1/\sqrt{\ell - \ell_n}$ as $\ell\to \ell_n = n/\left[1+\frac{a_0^2(1+\delta^2)}{2(1+\rho_\perp^2)}\right]$, the locations of the IPW harmonics. The underlying reason is that the condition $\zeta(\varphi)=n$ in the argument of the delta distribution in \eqref{Eq.2.25} becomes stationary (i.e., $\zeta'\left(\varphi\right)=0$), resulting in the fold-type caustic in the emitted spectrum \cite{Seipt2016a, Kharin2018, KravtsovCaustics1983, Fedotov2023}. In the following, we introduce a procedure to ``remedy'' the LMA probability rate of these divergencies by reintroducing finite bandwidth effects.

    \subsection{Derivation of the LMA$^+$}

    To obtain a divergence-free fully differential probability, we address a subtlety in the formal derivation of the LMA, namely the evaluation of $\theta$-integral in (\ref{Eq.2.11}) yielding the delta distribution in (\ref{Eq.2.16}). This delta distribution is accumulated, when one integrates over the whole range of the $\theta$ variable, including contributions even from the interference windows $\theta$ that do not formally satisfy the condition $\theta/\Delta \ll 1$. The way we resolve this inconsistency is by introducing a window function $\mathcal W (\theta)$ into expression \eqref{Eq.2.11}, with the purpose of excluding contributions from the $\theta$-regions, where the condition $\theta/\Delta\ll1$ is not fulfilled,
     \begin{equation}
    \frac{d\mathbb{R}_\mathrm{LMA^+}}{d\ell d^2\vec{\rho}_\bot} 
        = 
        -\frac{\alpha}{\pi^2}{A}\sum_{n=-\infty}^{+\infty} \mathbb{D}_n\left(\varphi\right) \int\limits_{-\infty}^{+\infty}\mathcal{W}\left(\theta\right)\exp\left[i\theta\left(\zeta\left(\varphi\right) - n\right)\right]d\theta  \,. 
     \label{Eq.3.1}
     \end{equation}
    This re-introduces a finite bandwidth of the background, that was neglected due to the local approximation. We denote this bandwidth-restored LMA as LMA$^+$.

    In principle, the window function $\mathcal W(\theta)$ could be chosen arbitrarily. Hereinafter, we model the envelope of the background with a Gaussian function $g\left(\varphi/\Delta\right) = \exp\left(-\varphi^2/2\Delta^2\right)$ and choose our window function to be Gaussian as well. 
    To find a suitable value for the width of the Gaussian window, we match the spectra of LMA$^+$ to the exact SFQED result in the weak field limit $a_0\ll1$. This prescription determines the window function as $\mathcal W (\theta) = \exp\left(- \theta^2/\Delta^2\right)$. We note that to achieve full correspondence of the spectra in the weak field limit, it is necessary to choose the window function in accordance with the envelope function.
    After performing the integration with respect to $\theta$ in Eq.~\eqref{Eq.3.1}, we obtain the LMA$^+$ rate as
    \begin{align}
    \frac{d\mathbb{R}_\mathrm{LMA^+}}{d\ell d^2\vec{\rho}_\bot}
    =
    -\frac{2\alpha\Delta}{\pi^{3/2}}{A}\sum_{n=1}^{+\infty} \mathbb{D}_n\left(\varphi\right) \exp\left[-\Delta^2\left(\zeta\left(\varphi\right)-n\right)^2\right]\,.
     \label{Eq.3.2}
    \end{align}

    In the standard LMA, the position of the harmonics is strictly defined by the delta distribution and the condition $\zeta\left(\varphi\right) = n$,
    yielding a relation between transferred momentum and laser phase as \cite{Seipt2013}
    \begin{align}
    \ell_n\left(\varphi\right) = \frac{n\left(1 + \vec{\rho}_\bot^2\right)}{1 + \vec{\rho}_\bot^2 + \frac{1+\delta^2}{2}a_0^2g^2\left(\varphi/\Delta\right)}.
    \label{harmonics_pos}
    \end{align}
    In the LMA$^+$, the delta distribution is replaced by Gaussian in Eq.~\eqref{Eq.3.2}. Thus, the harmonic lines are rather ``smeared'' in the vicinity of the original positions, Eq.~\eqref{harmonics_pos}, and the width of the harmonic lines is $\propto \Delta^{-1}$. The finite bandwidth contribution in the harmonics is the consequence of the finite temporal domain assigned to the $\theta$-integral via Gaussian window function in  Eq.~\eqref{Eq.3.1}. The absence of the delta distribution in (\ref{Eq.3.2}) also implies that $n$ could be zero or a negative integer in the LMA$^+$. However, since $\Delta \gg 1$, and the assigned harmonics bandwidth is small, the contributions from the zeroth and negative harmonics are negligible \cite{King2021}, so we will not discuss them any further.
    
     \begin{figure}[th]
    \centering
    \includegraphics[width=\textwidth]{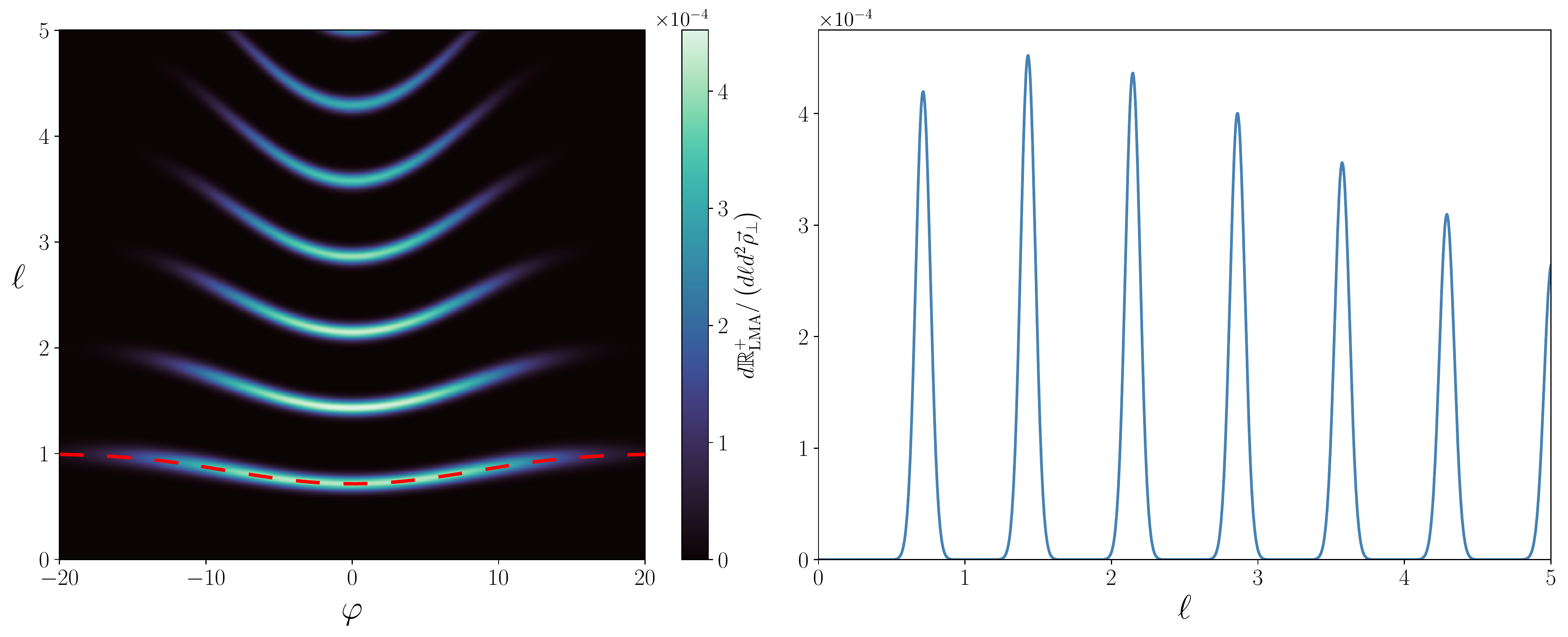}
    \caption{Fully differential $\mathrm{LMA}^+$ probability rate of NCS in a LP pulse as a function of $\ell$ and $\varphi$ as a contour plot (left).
    The location of the first harmonic's center, according to Eq.~\eqref{harmonics_pos}, is depicted by the red dashed curve.
    The right panel shows a lineout at $\varphi = 0$. The other parameters are: $a_0 = 2$, $\Delta = 10$, $\eta = 0.1$, $\vec{\rho}_\bot =\left(a_0, 0\right)$. \label{fig.1}
    }
    \end{figure}

    In Fig.~\ref{fig.1} we display the distribution of the fully differential probability rate as a function of the laser phase $\varphi$ and the transferred momentum $\ell$. This distribution tells us how the harmonic structure of the emitted spectrum is formed throughout the laser pulse. The red dashed curve in the left panel of Fig.~\ref{fig.1} designates the position of the first harmonic according to Eq.~\eqref{harmonics_pos}. Each harmonic exhibits phase-dependent red-shift that results into the broadening \cite{Seipt_chirp_2015} and the overlapping \cite{Seipt2011} of harmonics after integrating over the laser phase $\varphi$. The maximum red-shift takes place in the middle point ($\varphi=0$) of the laser pulse (see, Eq.~\eqref{harmonics_pos}), where the field intensity is the largest. We see that the higher harmonics are emitted closer to the pulse peak, which agrees with the previous theoretical studies \cite{Seipt2016}. To demonstrate the recovered finite bandwidth, we plot the lineout for $\varphi=0$ in the right panel of the Fig.~\ref{fig.1}. Due to the specific choice of the window function in \eqref{Eq.3.1}, the emitted harmonics have Gaussian profile. We note that for the standard LMA, harmonics lack any width at all and form a delta-comb, making probability rate distribution meaningless.

    \subsection{Numerical results for the fully differential LMA$^+$ probability}
    \label{sect:lma+numeric}
    {In this section, we show the numerical results for various observables and discuss the underlying physics, contrasting the LMA and LMA$^+$ with calculations of the NCS probability calculated by using the exact SFQED $S$-matrix elements \cite{Seipt2011, Boca2009, Krajewska2012}, and the angularly resolved LCFA \cite{Blackburn2020}}

\begin{figure}[ht]
    \centering
    \includegraphics[width=\textwidth]{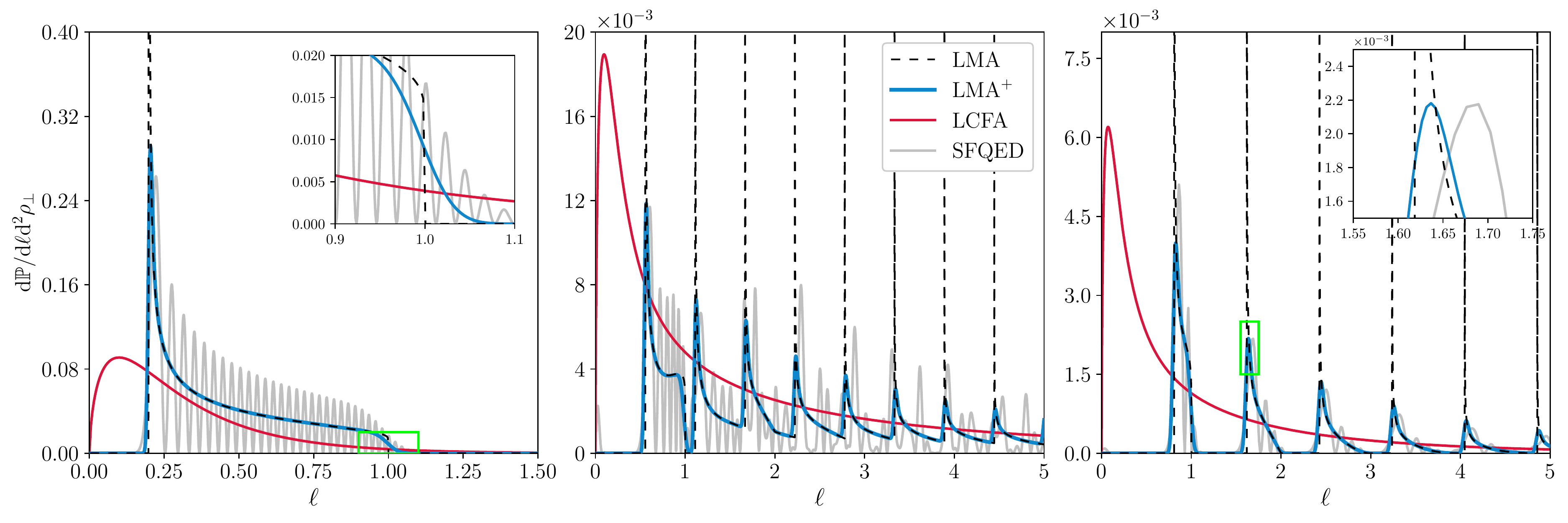}
    \caption{Fully differential NCS spectra in CP background plotted for the different values of the transverse momentum $\vec{\rho}_\bot$:  $\vec{\rho}_\bot=\left(0,0\right)$-left (on axis),  $\vec{\rho}_\bot=\left(a_0,0\right)$-center,  $\vec{\rho}_\bot=\left(2a_0,0\right)$-right. The classical nonlinear parameter $a_0=2$, pulse width $\Delta = 25$ and the initial electron energy parameter $\eta = 0.1$.
    \label{fig.2}}
    \end{figure}

    We start our investigation by comparing in Fig.~\ref{fig.2} the triple-differential \emph{probabilites} $d\mathbb P/d\ell d^2 \rho_\perp$ as function of $\ell$ for $a_0=2$ and $\Delta=25$, and for fixed values of $\vec \rho_\perp$ in each panel. In general, both the LMA and LMA$^+$ are reasonable approximations of the full SFQED spectra for these parameters. The LCFA, on the other hand, is not capable of correctly reproducing the harmonic spectra here for $a_0\sim O\left(1\right)$---it is outside the realm of its applicability.

    {Both the LMA and the LMA$^+$ correctly predict the harmonic structure of the NCS spectrum. For a finite pulse, the location of the $n$-th harmonic is approximately given by the condition \cite{Seipt2013}
    \begin{align}
    \ell_n(\varphi=0) = n \frac{1 + \vec{\rho}_\bot^2}{1 + \vec{\rho}_\bot^2 + a_0^2} \leq \ell \leq n \label{harm_range}%
    \end{align}
    with $\ell_n(\varphi)$ given in Eq.~\eqref{harmonics_pos}. In the LMA, this condition is \emph{strict}, while the SFQED result slightly spreads outside this region due to the finite background pulse bandwidth. This behavior can be seen particularly well in the inset in the left panel of Fig.~\ref{fig.2}. The LMA$^+$ correctly reproduces this behavior of the SFQED result. More crucially, the LMA diverges at the lower boundaries of the harmonic ranges, while both the SFQED and LMA$^+$ results stay finite. Apart from these differences at the endpoints of the harmonic ranges, the LMA$^+$ result agrees with the standard LMA.}

    {Neither the LMA nor the LMA$^+$ can reproduce the characteristic sub-harmonic structure of the full SFQED spectrum \cite{Seipt2011, Seipt2016}; they yield an average over these high-frequency oscillations. The sub-harmonic structure is due to long-range interference between the two indistinguishable events (e.g., emission of the photons of the same energy at the same angle in different points of a laser pulse), separated by the distance $\theta\sim\Delta$, and is inaccessible, if one adheres to the local approximation $\theta/\Delta\ll1$.}

    {We note that the positions of harmonics maxima in the LMA$^+$ have a slight offset compared to the SFQED result (see in particular the inset in the right panel of Fig.~\ref{fig.2}). This small discrepancy stems from our windowing procedure (\ref{Eq.3.1}). Due to the specific choice of the window function $\mathcal W \left(\theta\right)$, each harmonic has a Gaussian profile in the vicinity of the cutoff (see also the lineouts in Fig.~\ref{fig.1}). In fact, the correct shape of the harmonic profile is closely related to the Airy function due to the fold-type caustic structure of the emitted spectrum \cite{Narozhnyi1996, Kharin2016, Seipt2016a, Kharin2018}.
    }

    \begin{figure}[ht]
    \centering
    \includegraphics[width=\textwidth]{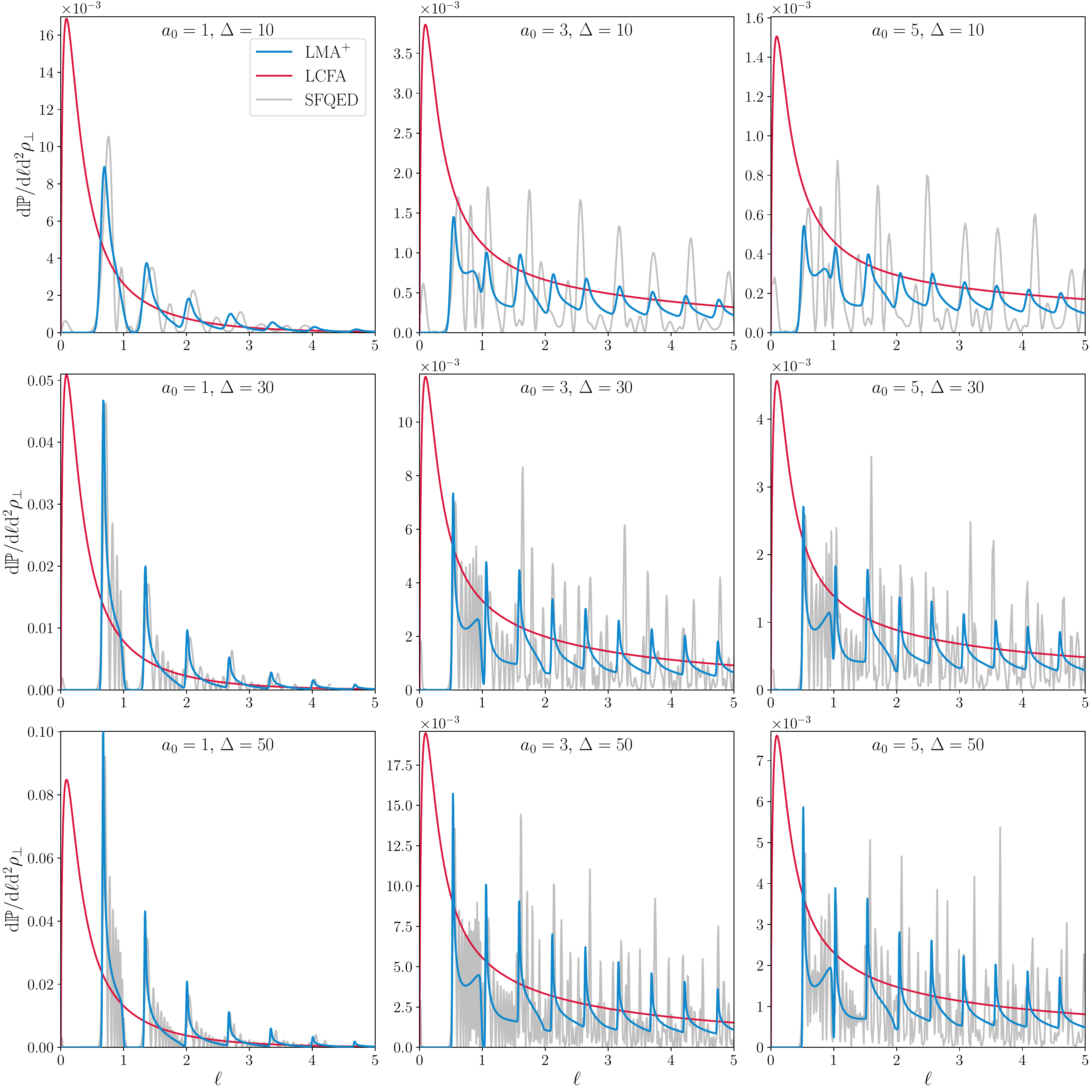}
    \caption{Fully differential NCS spectra in CP background for the different intensities and pulse durations. The transverse momentum is always chosen as $\vec{\rho}_\bot=\left(a_0,0\right)$ and the initial electron energy parameter $\eta = 0.1$.
        \label{fig.3}}
    \end{figure}
    
    {In Fig.~\ref{fig.3} we address how the LMA$^+$ works for the different laser strength $a_0$ and pulse durations $\Delta$.} We see that better agreement is achieved for the smaller $a_0$ and the longer pulses. With the growth of intensity, the overlap of harmonics and sub-harmonic structure result into the presence of additional sub-peaks that are even more pronounced, than the harmonics themselves. For smaller values of $a_0$, harmonics do not overlap, and the additional peaks are missing. Whereas, the longer the pulse, the better LMA$^+$ averages through the sub-harmonic structure, which is also true for the larger $a_0$.

    \begin{figure}[h!]
    \centering
    \includegraphics[width=\textwidth]{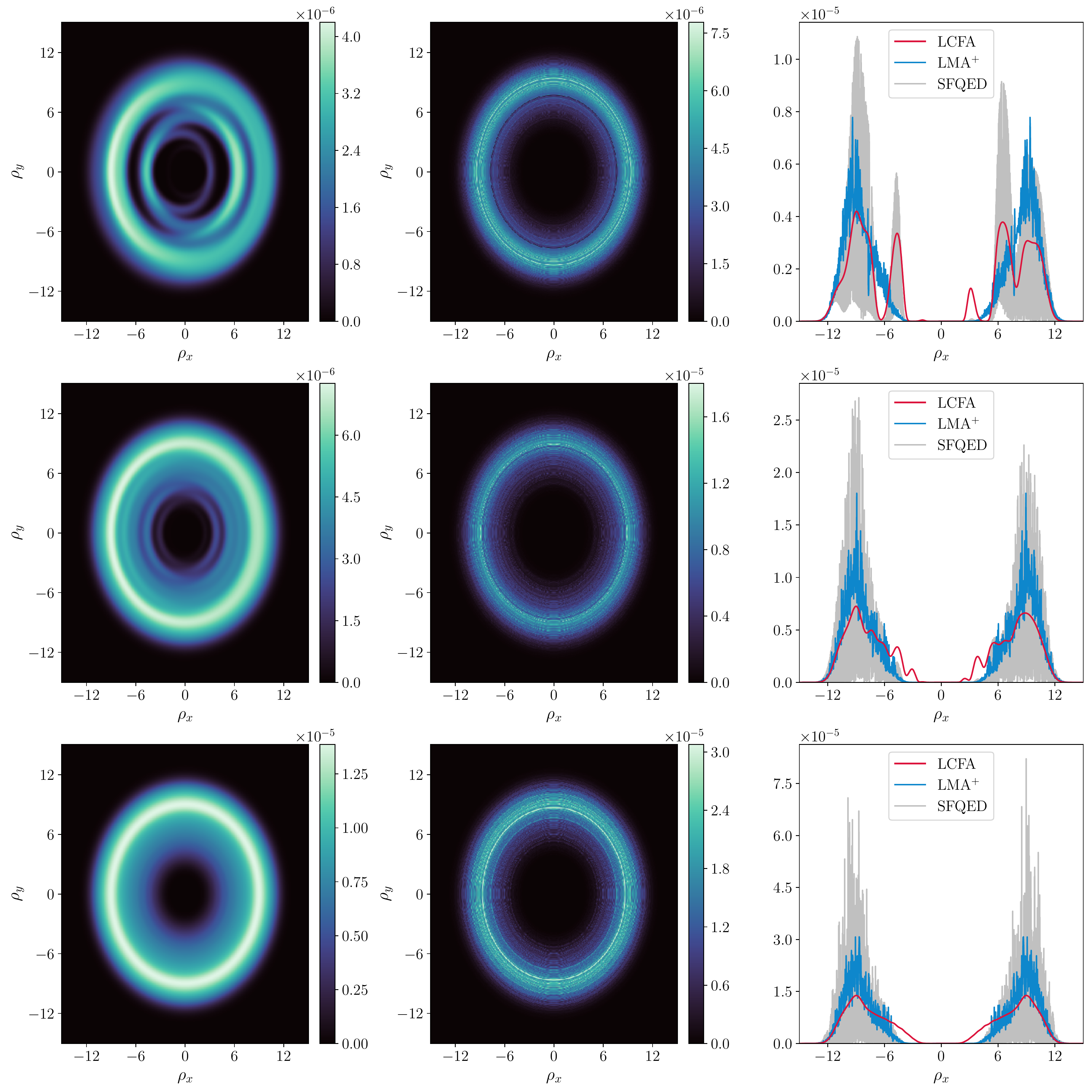}
    \caption{Fully differential NCS distributions in CP case for $a_0=10$, $\ell = 225$, $\eta = 0.1$ and different pulse durations: $\Delta = 10$ (top row), $\Delta = 20$ (middle row) and $\Delta = 40$ (bottom row). The left and right columns show LCFA and LMA$^+$ results, respectively. The right column demonstrates lineouts at $\rho_y = 0$ for both approximations and comparison with the full SFQED result.  
        \label{fig.4}}
    \end{figure}

    {Next, we investigate $d\mathbb P/d\ell d^2\rho_\perp$ as a function of $\vec \rho_\perp$ for fixed $\ell$. In Fig.~\ref{fig.4} we compare these angular distributions of the emitted radiation in the LMA$^+$ with results obtained in the LCFA and the exact SFQED calculations for the CP case. We note that this quantity cannot be easily plotted in the standard LMA due to the occurring delta distributions. The qualitative difference between the LCFA (first column) and LMA$^+$ (second column) is that the former predicts radiation beaming along the direction of the electron instantaneous momentum \cite{Blackburn2020, Blackburn2021}, which follows the four-potential vector trajectory in the transverse plane. Meanwhile, the latter manifests azimuthally symmetric distribution, due to the averaging over the cycle scale.} In the third column in Fig.~\ref{fig.4} we compare lineouts of the LCFA and LMA$^+$ results for $\rho_y=0$ with the exact SFQED calculations. We see that also the SFQED result has distinct peaks due to the radiation beaming just like the LCFA (top row), which is not seen in the LMA$^+$ result.
 
 The qualitative agreement is achieved between the LCFA, LMA$^+$ and SFQED only when the pulse is sufficiently long for a given intensity, so the emitted radiation from many cycles averages to a homogeneous pattern (see the last row in Fig.~\ref{fig.4}). To ensure this, we formulate an additional restriction on the applicability of the LMA$^+$. Given that the radiation beaming width in the $\left(\rho_x,\rho_y\right)$-plane is of the order of unity \cite{Blackburn2020}, we demand $||\vec{a}_\bot\left(\varphi+2\pi\right) - \vec{a}_\bot\left(\varphi\right)||\lesssim1$, so the two neighboring maxima of the background profile would be close enough, to guarantee overlap of the radiation due to the beaming. If this is not the case, the exact distribution will possess an additional structure, which LMA$^+$ is incapable to reproduce. {This yields the condition $\Delta \gtrsim 2\pi a_0$, that limits the applicability of the LMA$^+$ for large $a_0$ much stronger than the well known criterion $\Delta \gg1 $. Our results here show that for large $a_0$ the LCFA seems to yield better agreement with SFQED at finite $\Delta$, compared to the LMA$^+$. These findings need to be contrasted with the prediction that the (standard) LMA converges to the LCFA for $a_0\gg1$ for certain integrated rates \cite{Heinzl2020}.}

    \subsection{Analytic results for integrated LMA$^+$ rates}
    \label{sect:lma+analytic}

    {In the expression for the standard LMA rate, Eq.~\eqref{Eq.2.25}, the delta distribution allows for an immediate exact analytical integration with respect to one of the three dependent variables: laser phase $\varphi$, transferred momentum $\ell$ or absolute value of the transverse momentum $|\vec{\rho}_\bot|$. In turn, for the LMA$^+$ (\ref{Eq.3.2}) this is not so straightforward. Nonetheless, we may perform the integration approximately, due to the huge damping factor $\sim\Delta^2$ in the exponent of Eq.~\eqref{Eq.3.2}. Below, in subsections \ref{subsecD} and \ref{subsecC} we demonstrate how to approximately perform analytically the integrations with respect to laser phase $\varphi$ and the transverse momentum magnitude $|\vec{\rho}_{\bot}|$, respectively.}

 \subsubsection{Integration over the laser phase $\varphi$}
    \label{subsecD}

Here we show how to obtain fully differential probability from the probability rate (\ref{Eq.3.2}), performing approximately integration over the laser phase $\varphi$.\todoNL{Should we put here explicit expression for the integral that we evaluate?} The integrand, Eq.~\eqref{Eq.3.2}, is suppressed, unless $\left(\zeta\left(\varphi\right)-n\right)\lesssim1/\Delta \ll 1$. Thus, the main contributions to the $\varphi$ integral come from the vicinities of zeros of $\zeta\left(\varphi\right)-n$. Since $\zeta\left(\varphi\right)-n$ is an even function of $\varphi$, it has two zeros, which lay symmetrically with respect to the origin
\begin{equation}
\varphi_* = \pm\Delta g^{-1}\left(\sqrt{\frac{2\left(1+\vec{\rho}_\bot^2\right)}{a_0^2\left(1+\delta^2\right)}\left(\frac{n}{\ell}-1\right)}\right).
    \label{Eq.3.3}
\end{equation}
So, we expand $\zeta\left(\varphi\right) - n$ in the vicinity of $\varphi_*$ up to the fourth order
 \begin{multline}      \label{Eq.3.4}
\left(\zeta ( \varphi ) - n\right)^2
 =
 \left[\zeta'\left(\varphi_*\right)\right]^2\left(\varphi-\varphi_*\right)^2 +\zeta'\left(\varphi_*\right)\zeta''\left(\varphi_*\right)\left(\varphi - \varphi_*\right)^3 \\
 + \left[\left(\frac{\zeta''\left(\varphi_*\right)}{2}\right)^2 + \frac{\zeta'\left(\varphi_*\right)\zeta'''\left(\varphi_*\right)}{3}\right]\left(\varphi-\varphi_*\right)^4 + O\left[\left(\varphi-\varphi_*\right)^5\right].
 \end{multline}
 
The inclusion of the fourth order term is necessary to ensure the finite result. When the two zeros coalesce $\varphi_* = 0$, the first derivative $\zeta'\left(\varphi_*\right)\propto g'\left(\varphi_*/\Delta\right)$ vanishes, and the correct behavior is controlled by the fourth term in (\ref{Eq.3.4}).
Moreover, we argue that the second term in the square brackets in (\ref{Eq.3.4}) may be also neglected, since the overall contribution of the fourth order term is important only when $\zeta'\left(\varphi_*\right) = 0$. Otherwise, the dominant role is played by the lower order terms (by the same token, we neglect the term $\propto \left(g'\right)^2$ in $\zeta''\left(\varphi_*\right)$, since it plays no role in the regularization as well.). Thus, we can write  

 \begin{equation}
\left(\zeta\left(\varphi\right)-n\right)^2\approx\left[\zeta'\left(\varphi_*\right)\left(\varphi - \varphi_*\right) + \frac{\zeta''\left(\varphi_*\right)}{2}\left(\varphi-\varphi_*\right)^2\right]^2.
    \label{Eq.3.5}
\end{equation}
Substituting (\ref{Eq.3.5}) into (\ref{Eq.3.2}), approximating the rest integrand by its value in $\varphi_*$ and shifting integration variable $\varphi - \varphi_* = t$, we may write

  \begin{equation}
 \frac{d\mathbb{P}_\mathrm{LMA^+}}{d\ell d^2\vec{\rho}_\bot}\approx-\frac{2\alpha\Delta}{\pi^{3/2}}{A}\sum_{n=1}^{+\infty} \mathbb{D}_n\left(\varphi_*\right)\int\limits_{-\infty}^{+\infty}dt \exp\left[-\Delta^2\left(\zeta'\left(\varphi_*\right)t + \frac{\zeta''\left(\varphi_*\right)}{2}t^2\right)^2\right].
     \label{Eq.3.6}
 \end{equation}
The integral over $t$ may be expressed via the modified Bessel functions of the first kind \cite{Watson1966} and the final analitic expression for the LMA$^+$ NCS probability is given by
    \begin{equation}
 \frac{d\mathbb{P}_\mathrm{LMA^+}}{d\ell d^2\vec{\rho}_\bot}\approx-\frac{\alpha\Delta}{\sqrt{\pi}}{A}\sum_{n=1}^{+\infty} \mathbb{D}_n\left(\varphi_*\right)\left|\frac{\zeta'\left(\varphi_*\right)}{\zeta''\left(\varphi_*\right)}\right|\left[I_{\frac{1}{4}}\left(\Delta^2\varrho^2\right)+I_{-\frac{1}{4}}\left(\Delta^2\varrho^2\right)\right]\exp\left(-\Delta^2\varrho^2\right),
     \label{Eq.3.7}
 \end{equation}
 where $\varrho ^2= \left[\zeta'\left(\varphi_*\right)\right]^4/8\left[\zeta''\left(\varphi_*\right)\right]^2$.

 We note that one can recover the standard LMA fully differential probability,
     \begin{equation}
 \frac{d\mathbb{P}_\mathrm{LMA}}{d\ell d^2\vec{\rho}_\bot}=-\frac{4\alpha}{\pi}{A}\sum_{n=1}^{+\infty} \frac{\mathbb{D}_n\left(\varphi_*\right)}{|\zeta'\left(\varphi_*\right)|},
     \label{Eq.3.11}
 \end{equation}
 from (\ref{Eq.3.7}), as the first term in the asymptotic expansion for $\Delta\gg1$, given the expression for the modified Bessel function of the first kind for the large argument \cite{Watson1966}: $I_\nu\left(x\right)\sim e^x/\sqrt{2\pi x}$ as $x\to+\infty$.

 The advantage of the result (\ref{Eq.3.7}) is that in contrast to the expression (\ref{Eq.3.11}), it stays finite for the vanishing $\zeta'\left(\varphi_*\right)$. Also, when the $\zeta''\left(\varphi_*\right)$ tends to zero ($\varrho \to\infty$), the finiteness of (\ref{Eq.3.7}) is ensured by the leading asymptotic term of $I_{\pm1/4}\left(\Delta^2\varrho^2\right)$. To the best of our knowledge, the expression (\ref{Eq.3.7}) is the first divergence-free result for the fully differential NCS probability. But, we stress that (\ref{Eq.3.7}) is not of the general nature, since it was obtained for the specific choice of the window function.

 \begin{figure}[ht]
\centering
 \includegraphics[width=\textwidth]{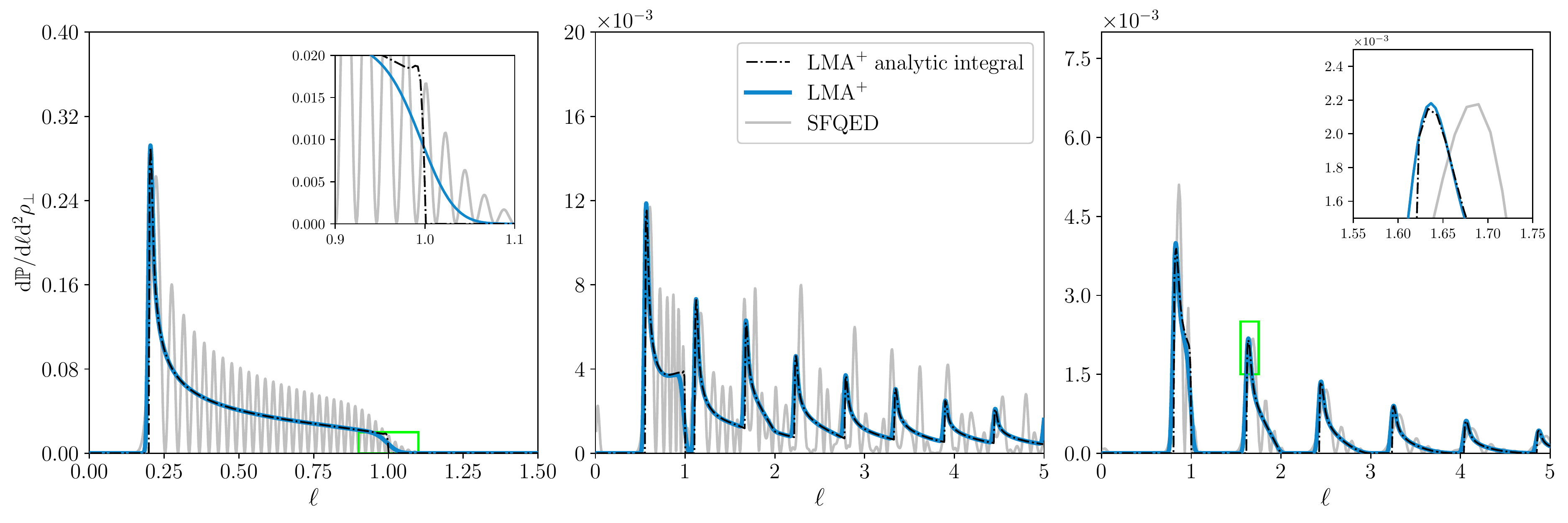}
\caption{Fully differential NCS spectra in CP background plotted for the different values of the transverse momentum $\vec{\rho}_\bot$:  $\vec{\rho}_\bot=\left(0,0\right)$-left (on axis),  $\vec{\rho}_\bot=\left(a_0,0\right)$-center,  $\vec{\rho}_\bot=\left(2a_0,0\right)$-right. The classical nonlinear parameter $a_0=2$, pulse width $\Delta = 25$ and the initial electron energy parameter $\eta = 0.1$.
    \label{fig.6}}
\end{figure}

In Fig.~\ref{fig.6} we now compare the result (\ref{Eq.3.7}) with curves obtained by the numerical integration of (\ref{Eq.3.2}) with respect to $\varphi$ and full SFQED result. We see that Eq.~\eqref{Eq.3.7} provides finite result at the left edge of the harmonics range, where standard LMA is divergent (see, also Fig.~\ref{fig.2}). However, due to approximations performed in the integration over $\varphi$, the harmonics spreading outside the allowed range, Eq~\eqref{harm_range}, disappeared. The expression (\ref{Eq.3.7}) predicts the abrupt edges (see inset in Fig.~\ref{fig.6}) for harmonics, likewise the standard LMA. Apart from that, approximation (\ref{Eq.3.7}) is in a good agreement with the numerically integrated LMA$^+$ result. This shows that the analytical result \eqref{Eq.3.7} manages to keep the finite bandwidth effects, where needed most: to shield the divergencies.

    \subsubsection{Integration over the transverse momentum $\vec \rho_\perp$}
    \label{subsecC}

Applying the same reasoning as above, we can approximately integrate the LMA$^+$ probability rate \eqref{Eq.3.2} over the transverse momentum magnitude $|\vec{\rho_\bot}|$. Further, we consider the CP case, since it also permits analytical integration over the azimuthal angle $\vartheta$ in the transverse momentum plane ($d^2\vec{\rho}_\bot =|\vec{\rho}_\bot|  d|\vec{\rho}_\bot|d\vartheta$) due to the symmetry of the background. 

To proceed, we make a change of variables in (\ref{Eq.3.2}) and (\ref{Eq.2.26}) and go from the transferred momentum $\ell$ to the light-cone momentum fraction $s = \kappa\cdot k/\kappa\cdot p= 2\eta\ell/\left(1 + \vec{\rho}^2_\bot + 2\eta\ell\right)$, and use that $d\ell/\ell = ds/s\left(1 - s\right)$. Doing so, (\ref{Eq.3.2}) turns into: 

   \begin{equation}
 \frac{d\mathbb{R}_\mathrm{LMA^+}}{ds}=-\frac{\alpha\Delta}{2\eta^2\pi^{3/2}}\frac{s}{1-s}\int\limits_{\mathbb{R}^2}d^2\vec{\rho}_\bot\sum_{n=1}^{+\infty} \mathbb{D}_n\left(\vec{\rho}^2_\bot\right) \exp\left[-\Delta^2\left(\zeta\left(\vec{\rho}_\bot^2\right)-n\right)^2\right],
     \label{Eq.3.12}
 \end{equation}

 \begin{equation}
 \mathbb{D}_n\left(\vec{\rho}^2_\bot\right) = J_n^2\left(x\right) + a_0^2g^2\left(\frac{\varphi}{\Delta}\right)\left[\frac{1}{2}+\frac{s^2}{4\left(1-s\right)}\right]\left(2J_n^2-J^2_{n+1}-J_{n-1}^2\right),
     \label{Eq.3.13}
 \end{equation}

 \begin{equation}
x=\frac{s}{\eta\left(1-s\right)}|\vec{\rho}_\bot|a_0g\left(\frac{\varphi}{\Delta}\right).
     \label{Eq.3.14}
 \end{equation}
Exploiting the azimuthal symmetry for integration over $\vartheta$ and introducing a new notation $r = \vec{\rho}_\bot^2$, we arrive at

    \begin{equation}
 \frac{d\mathbb{R}_\mathrm{LMA^+}}{ds}=-\frac{\alpha\Delta}{2\eta^2\sqrt{\pi}}\frac{s}{1-s}\int\limits_0^{+\infty}dr\sum_{n=1}^{+\infty} \mathbb{D}_n\left(r\right) \exp\left[-\left(\frac{\Delta s}{2\eta\left(1-s\right)}\right)^2\left(r-r_*\right)^2\right],
     \label{Eq.3.15}
 \end{equation}
 
 \begin{equation}
 r_* = \frac{2\eta n\left(1-s\right)}{s}\left[1 - \frac{s}{\bar{s}\left(1-s\right)}\right],\quad \bar{s} = \frac{2\eta n}{1 + a^2\left(\varphi/\Delta\right)}.
     \label{Eq.3.16}
 \end{equation}
The main contribution to the integral (\ref{Eq.3.15}) comes from the region, where $r-r_*\lesssim1/\Delta$. Hence, we may approximate \eqref{Eq.3.15} as
     \begin{equation}
 \frac{d\mathbb{R}_\mathrm{LMA^+}}{ds}\approx-\frac{\alpha\Delta}{2\eta^2\sqrt{\pi}}\frac{s}{1-s}\sum_{n=1}^{+\infty} \mathbb{D}_n\left(r_*\right)\int\limits_0^{+\infty}dr \exp\left[-\left(\frac{\Delta s}{2\eta\left(1-s\right)}\right)^2\left(r-r_*\right)^2\right].
     \label{Eq.3.18}
 \end{equation}
 Evaluating the integral \eqref{Eq.3.18} with respect to $r$, we obtain a complementary error function \cite{Olver2010}, and the final result reads as 
      \begin{equation}
 \frac{d\mathbb{R}_\mathrm{LMA^+}}{ds}\approx-\frac{\alpha}{\eta}\sum_{n=1}^{+\infty} \mathbb{D}_n\left(r_*\right)\left[1 - \frac{1}{2}\mathrm{Erfc}\left(n\Delta\left[1 -\frac{s}{\bar{s}\left(1-s\right)}\right]\right)\right],
     \label{Eq.3.19}
 \end{equation}
with the argument of the Bessel functions in (\ref{Eq.3.13}) being now
 \begin{equation}
x=\frac{2na\left(\varphi/\Delta\right)}{\sqrt{1+a^2\left(\varphi/\Delta\right)}}\sqrt{\frac{s}{\bar{s}\left(1-s\right)}\left(1-\frac{s}{\bar{s}\left(1 -s\right)}\right)}.
     \label{Eq.3.20}
 \end{equation}

 The first term in the square brackets in (\ref{Eq.3.19}) gives the standard LMA expression \cite{Heinzl2020}. In turn, the complementary error function corresponds to the finite bandwidth corrections. When the value of $s$ is far from the harmonic edge $s_n = \bar{s}/\left(1 + \bar{s}\right) = 2\eta n/\left(1 + 2\eta n + a^2\left(\varphi/\Delta\right)\right)$, the argument of the complementary error function $\sim n\Delta\gg1$, and we have a strong suppression of the second term in (\ref{Eq.3.19}), since $\mathrm{Erfc}\left(x\right)\sim e^{-x^2}/x\sqrt{\pi}$ for the large positive $x$ \cite{Olver2010}. In other words, the standard LMA and LMA$^+$ probability rates coincide far from the harmonics positions. But, when $s$ approaches the harmonic edge $s_n$ the argument of the complementary error function in (\ref{Eq.3.19}) becomes of the order of unity, and the second term damps the probability rate's magnitude. 

Another significant difference between the LMA and LMA$^+$ is the allowed range of change for the $s$ variable. The standard LMA predicts that, for each individual harmonic $n$, $s$ is confined $0\leq s\leq s_n$ \cite{Heinzl2020}. This condition stems from the delta distribution (\ref{Eq.2.25}) and manifests zero bandwidth of the monochromatic model.
However, it is not the case for the LMA$^+$. The assigned finite bandwidth leads to the softening of the restriction on the valid range for the $s$ variable. 

When $s> s_n$, the sign of the complementary error function argument changes, and since $\mathrm{Erfc}\left(-x\right) = 2 - \mathrm{Erfc}\left(x\right)$, we get

\begin{equation}
 \frac{d\mathbb{R}_\mathrm{LMA^+}}{ds}\approx\frac{\alpha}{2\eta}\sum_{n=1}^{+\infty} \left(-1\right)^n\mathbb{D}_n\left(r_*\right)\mathrm{Erfc}\left(n\Delta\left[\frac{s}{\bar{s}\left(1-s\right)} - 1\right]\right),
     \label{Eq.3.21}
 \end{equation}
where
  \begin{equation}
 \mathbb{D}_n\left(r_*\right) = I_n^2\left(\tilde{x}\right) + a^2\left(\frac{\varphi}{\Delta}\right)\left[\frac{1}{2}+\frac{s^2}{4\left(1-s\right)}\right]\left(2I_n^2+I^2_{n+1}+I_{n-1}^2\right),
     \label{Eq.3.22}
 \end{equation}

  \begin{equation}
 \tilde{x}=\frac{2na\left(\varphi/\Delta\right)}{\sqrt{1+a^2\left(\varphi/\Delta\right)}}\sqrt{\frac{s}{\bar{s}\left(1-s\right)}\left(\frac{s}{\bar{s}\left(1 -s\right)}-1\right)}.
     \label{Eq.3.23}
 \end{equation}
To obtain (\ref{Eq.3.22}), we employed the relation: $J_n\left(ix\right) = e^{i\pi n/2}I_n\left(x\right)$ \cite{Watson1966}. We see from (\ref{Eq.3.21}) and (\ref{Eq.3.22}) that for $s\to1$ the modified Bessel functions tend to infinity, however, this divergence is suppressed by the faster decay of the complementary error function, so the overall expression stays finite and, as we will see, predicts reasonable behavior in the high-energy region of the emitted spectrum.

\begin{figure}[ht]
\centering
 \includegraphics[width=0.495\textwidth]{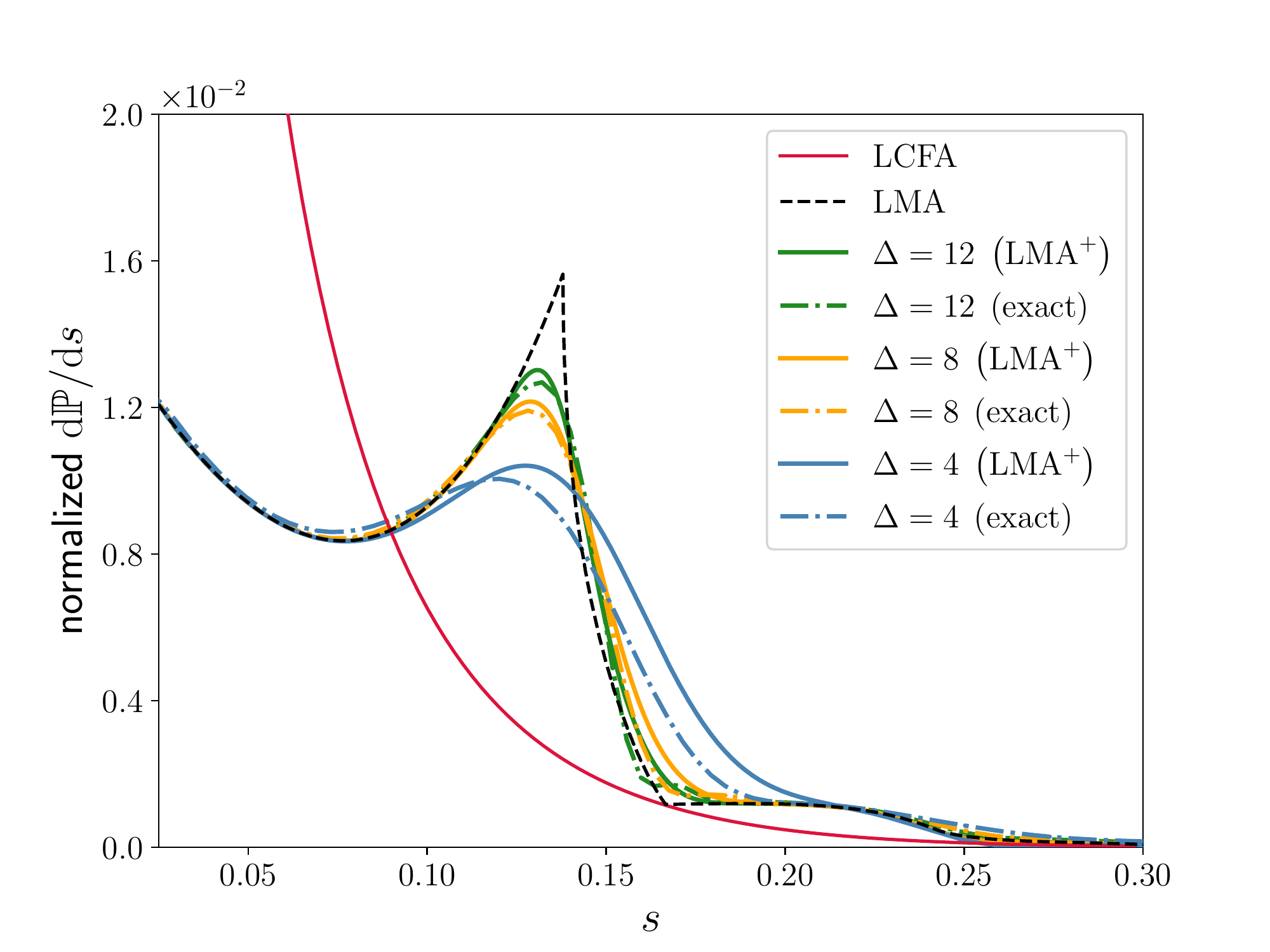}
  \includegraphics[width=0.495\textwidth]{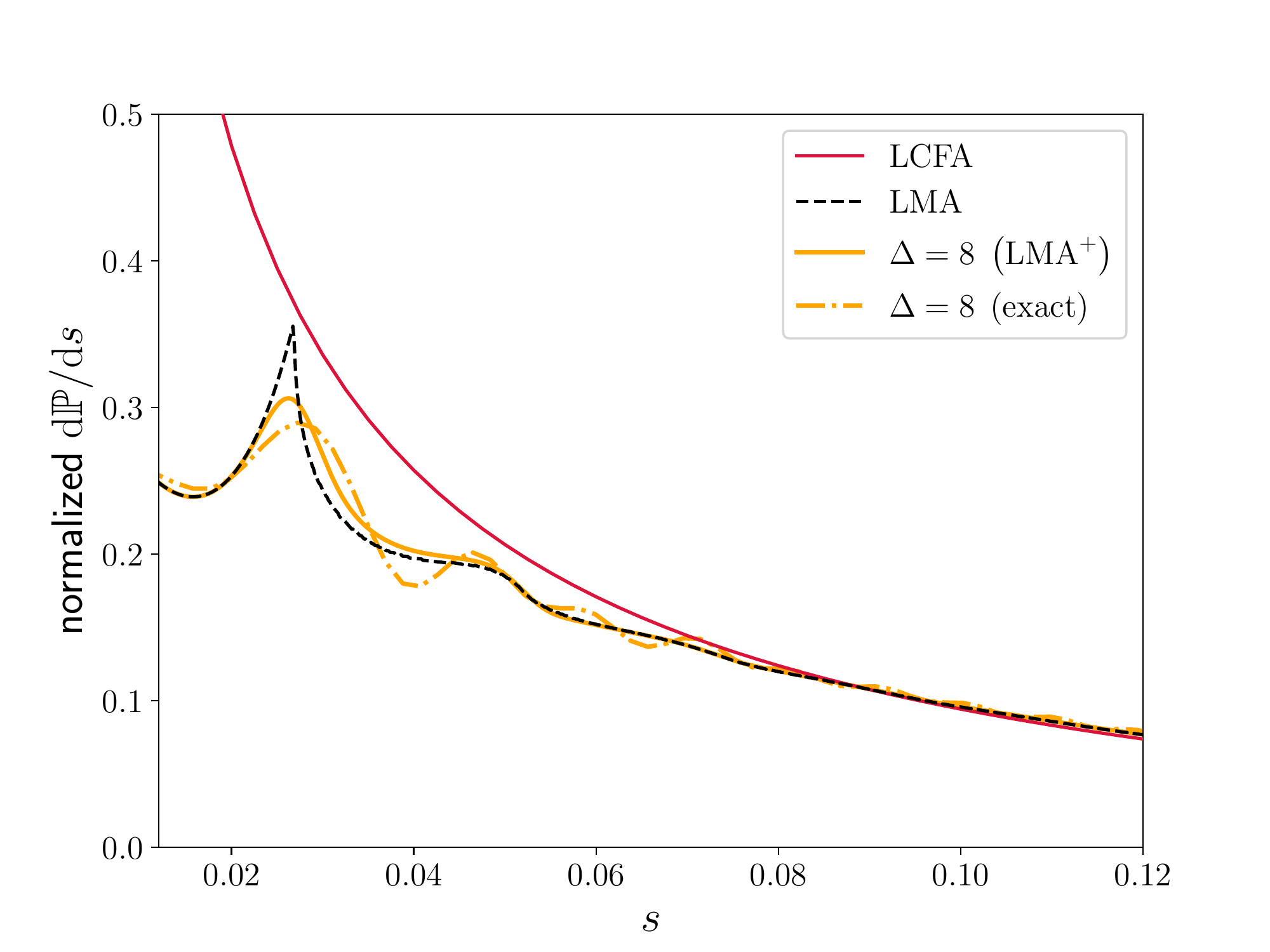}
\caption{Angular-integrated spectra of NCS in CP background plotted for $a_0 =0.5$ (left), $a_0=2.5$ (right). Each curve is normalized by the total pulse duration $\Delta$. The initial electron energy parameter $\eta = 0.1$.
    \label{fig.5}}
\end{figure}

In Fig.~\ref{fig.5} we compare the exact SFQED results with LMA$^+$ given by the Eqs.~\eqref{Eq.3.19} and (\ref{Eq.3.21}), each curve is normalized by the total pulse duration $\Delta$. We also put the standard LMA and LCFA for the completeness. As we can see, the standard LMA is insensitive to the bandwidth features, such as spreading of the harmonics and damping of the harmonics peaks. LMA$^+$ reproduces these features to a certain extent: the harmonics magnitudes are substantially better approximated even for the relatively short pulses ($\Delta\not\gg1$), where SVEA is not valid anymore. Presumably, inclusion of the higher order gradients of the envelope, may help to achieve a better agreement with the exact calculations.

    \section{Summary}\label{sec:summary}

    {In the present paper, we revisited the derivation of the probability rates in the locally monochromatic approximation (LMA). Taking the probability of the nonlinear Compton scattering (NCS) as a starting point, we employed the slowly varying envelope approximation (SVEA) along with the local approximation ($\theta/\Delta\ll1$) to obtain the LMA proto-rate. The proper LMA probability rates were then obtained by cycle-averaging the proto-rate. The procedure works similarly for arbitrary polarization of a plane wave background.
    Our expressions for obtained LMA rates are in agreement with the ones from the literature, but do not require additional numerical arguments. Therefore, our novel approach emphasizes the key concepts behind the LMA: (i) the separation of time scales by means of the SVEA, (ii) neglect of interference effects on the scales of a full pulse length by employing the local approximation $\theta/\Delta\ll1$, and (iii) cycle-averaging over the fast carrier component to arrive at a positively semi-definite probability rate.}

    {We were able to restore the bandwidth effects into the LMA, which we call LMA$^+$. This is achieved in our new derivation of the LMA by introducing a window function in the proto-rate. By reintroducing finite bandwidth into the LMA probability rate, we find that the fully triple-differential probability within the LMA$^+$ stays finite, in contrast to the analog expression in the standard LMA. Our numerical comparison shows that the LMA$^+$ is able to reproduce the spreading of the emitted harmonics edges, similarly to the exact strong-field QED (SFQED) result, in contrast to the LMA. 
    We also investigate angular distribution of emitted radiation at fixed transferred momentum $\ell$. Considering especially the regime of $a_0\gg1$ and $\Delta\gg1$, we compared the angular distributions obtained with the LMA$^+$, the LCFA and the exact SFQED calculations. If $\Delta/a_0\ll2\pi$ the exact SFQED distribution manifests features associated with the sub-cycle structure of the pulse via the radiation beaming. While LCFA is capable to partially reproduce these features, the LMA$^+$ completely lacks them, due to the cycle-averaging procedure that hides all sub-cycle information. This yields a stronger constraint on the applicability of the LMA$^+$ as $\Delta/a_0\gtrsim2\pi$. If this criterion is violated, the LCFA was found to better agree with the full SFQED result than the LMA$^+$.}

    {Finally, we obtained new approximate analytical expression for integrated LMA$^+$ rates. First, we found expressions for the fully differential probability by integrating the LMA$^+$ over the laser phase $\varphi$. This analytical result correctly regularizes the LMA spectrum by removing the divergences in the standard LMA. However, due to the approximations made, it does not include the spreading of the harmonics due to the background's bandwidth. Second, we derived an approximate analytical expression for the angular-integrated probability rate in terms of modified Bessel functions, which accounts for the softening of the harmonic edges and damping of the harmonics peaks. Since the exact SFQED spectrum manifests the same features, we propose the LMA$^+$ probability rate as a more suitable candidate for the simulations of the laser-particle collisions in the moderate intensity regime $a_0\sim1$.}

\appendix

\section{Series representation for the floating average}
\label{sec:app.A}

We start with the definition of a floating average (\ref{Eq.2.4}) and perform the shift of the integration variable $t = \eta + \varphi$:
\begin{align}
\langle F\rangle\left(\varphi,\theta\right) = \frac{1}{\theta}\int\limits_{\varphi-\theta/2}^{\varphi+\theta/2}F\left(t\right)dt = \frac{1}{\theta}\int\limits_{-\theta/2}^{\theta/2}F\left(\eta+\varphi\right)d\eta.
    \label{A.1}
\end{align}
Treating $\langle F\rangle\left(\varphi,\theta\right)$ as a function of two variables, we write its Fourier transform with respect to the average laser phase $\varphi$
\begin{align}
\mathcal{F}\left(\lambda,\theta\right) = \frac{1}{2\pi}\int\limits_{-\infty}^{\infty} \langle F\rangle\left(\varphi,\theta\right)\exp\left(-i\lambda\varphi\right)d\varphi,
    \label{A.2}
\end{align}
and then substitute (\ref{A.1}) into the right-hand side of (\ref{A.2}), changing the integration order:
\begin{align}
\mathcal{F}\left(\lambda,\theta\right) =  \frac{1}{2\pi\theta}\int\limits_{-\theta/2}^{\theta/2}d\eta\int\limits_{-\infty}^{\infty}F\left(\eta+\varphi\right)e^{-i\lambda\varphi}d\varphi.
    \label{A.3}
\end{align}
For the inner integral with respect to $\varphi$ we perform a change of variables once again $\mu = \varphi + \eta$ and rearrange the integrands: 
\begin{align}
\mathcal{F}\left(\lambda,\theta\right) = \frac{1}{2\pi\theta}\int\limits_{-\theta/2}^{\theta/2}d\eta e^{i\lambda\eta}\int\limits_{-\infty}^{\infty}F\left(\mu\right)e^{-i\lambda\mu}d\mu = \frac{\tilde{F}\left(\lambda\right)}{\theta}\int\limits_{-\theta/2}^{\theta/2}d\eta e^{i\lambda\eta}.
    \label{A.4}
\end{align}
Here we also introduced the Fourier transform $\tilde{F}\left(\lambda\right)$ of the function $F\left(t\right)$. The rest integral with respect to $\eta$ can be done, giving us
\begin{align}
\mathcal{F}\left(\lambda,\theta\right) = {\tilde{F}\left(\lambda\right)}\mathrm{sinc}\left(\frac{\lambda\theta}{2}\right),
    \label{A.5}
\end{align}
where $\mathrm{sinc}\left(x\right)=\sin x/x$. The next step is to return to the original (\ref{A.1}) by means of the inverse Fourier transform of (\ref{A.5})
\begin{align}
\langle F\rangle\left(\varphi,\theta\right) = \int\limits_{-\infty}^\infty{\tilde{F}\left(\lambda\right)}\mathrm{sinc}\left(\frac{\lambda\theta}{2}\right)e^{i\lambda\varphi}d\lambda.
    \label{A.6}
\end{align}
If we replace $\mathrm{sinc}\left(\lambda\theta/2\right)$ with its Taylor series and change the order of integration and summation, we can write down the following expression: 
\begin{align}
\langle F\rangle\left(\varphi,\theta\right) =\sum_{n=0}^\infty \frac{\left(-1\right)^n}{\left(2n + 1\right)!}\left(\frac{\theta}{2}\right)^{2n}\int\limits_{-\infty}^\infty\lambda^{2n}\tilde{F}\left(\lambda\right)e^{i\lambda\varphi}d\lambda.
    \label{A.7}
\end{align}
The factor $\left(-1\right)^n\lambda^{2n}$ appears after taking $2n^{\mathrm{th}}$ derivative of the integrand with respect to the $\varphi$, so we can write
\begin{align}
\langle F\rangle\left(\varphi,\theta\right) =\sum_{n=0}^\infty \frac{1}{\left(2n + 1\right)!}\left(\frac{\theta}{2}\right)^{2n}\frac{d^{2n}}{d\varphi^{2n}}\int\limits_{-\infty}^\infty\tilde{F}\left(\lambda\right)e^{i\lambda\varphi}d\lambda.
    \label{A.8}
\end{align}
Eventually, we substitute the Fourier transform of the image $\tilde{F}\left(\lambda\right)$ with its original $F\left(\varphi\right)$ and obtain the final result:
\begin{align}
\langle F\rangle\left(\varphi,\theta\right) =\sum_{n=0}^\infty \frac{1}{\left(2n + 1\right)!}\left(\frac{\theta}{2}\right)^{2n}\frac{d^{2n}F\left(\varphi\right)}{d\varphi^{2n}}.
    \label{A.9}
\end{align}

\section{Cycle-averaging procedure}
\label{sec:app.B}

In this appendix, we show how to deal with the integrals that arise due to the cycle-averaging in (\ref{Eq.2.24}). For the CP case ($\delta = \pm 1$), the parameter $y$ vanishes (see, Eq.~\eqref{Eq.2.15}) and the generalized Bessel function simplifies to the standard Bessel function of the first kind (see, Eq.~\eqref{Eq.2.14}). Thus, the integral that we need to evaluate is 
\begin{align}
\int\limits_{\varphi-\pi}^{\varphi+\pi}J_n\left(2x\cos\varphi'\right)d\varphi' = 2\int\limits_{0}^{\pi} J_{2n}\left(2x\cos\varphi'\right)d\varphi'.
    \label{Eq.2.17}
\end{align}

In (\ref{Eq.2.17}) we beforehand shifted the argument $\varphi' - \vartheta \to\varphi'$ and then exploited properties of periodic functions along with the symmetries of the Bessel functions. In doing so, we assumed that the envelope ($x\sim g\left(\varphi'/\Delta\right)\approx g\left(\varphi/\Delta\right)$) stays constant on the cycle scale due to the SVEA. The integral on the right-hand side of (\ref{Eq.2.17}) is known as the Neumann integral \cite{Watson1966}:
\begin{align}
 \int\limits_{0}^{\pi} J_{2n}\left(2x\cos\varphi'\right)d\varphi' =  \pi J^2_n\left(x\right).
    \label{Eq.2.18}
\end{align}

In the LP case ($\delta = 0$), we deal with the two integrals of the generalized Bessel functions
\begin{align}
 \int\limits_{\varphi-\pi}^{\varphi+\pi}\mathcal{J}_{n}\left(2x\cos\varphi',2y\cos2\varphi'\right)d\varphi' 
 & =  2\int\limits_{0}^{\pi}\mathcal{J}_{2n}\left(2x\cos\varphi',2y\cos2\varphi'\right)d\varphi' \,,     \label{Eq.2.19} \\
\int\limits_{\varphi-\pi}^{\varphi+\pi}\cos2\varphi'\mathcal{J}_{n}\left(2x\cos\varphi',2y\cos2\varphi'\right) d\varphi' 
& = 2\int\limits_{0}^{\pi}\cos2\varphi'\mathcal{J}_{2n}\left(2x\cos\varphi',2y\cos2\varphi'\right) d\varphi' \,,
    \label{Eq.2.20}
\end{align}
where to get the right-hand side, we once again exploited properties of the periodic functions and symmetries of the Bessel function. The integrals (\ref{Eq.2.19}) and (\ref{Eq.2.20}) are the generalizations of the Neumann integral (\ref{Eq.2.18}), but since they are not well-known in the literature, we provide their evaluation here.

We adapt the strategy from Watson \cite{Watson1966} for the standard Neumann integral and for that we need the integral representation of the generalized Bessel function \cite{Dattoli1990}
\begin{align}
\mathcal{J}_n\left(x,y\right) = \frac{1}{2\pi}\int\limits_{-\pi}^\pi d\varphi\exp\left(ix\sin\varphi +iy\sin2\varphi - in\varphi\right).
    \label{B.1}
\end{align}
Then we consider the product of two generalized Bessel functions with different indices 
\begin{align}
\mathcal{J}_{n-m}\left(x,y\right)\mathcal{J}_{n+m}\left(x,y\right)&= \nonumber \\
&\hspace{-3cm}\frac{1}{4\pi^2}\int\limits_{-\pi}^\pi\int\limits_{-\pi}^\pi d\varphi d\theta\exp\left[-in\left(\varphi +\theta\right) + im\left(\varphi - \theta\right) + ix\left(\sin\varphi + \sin\theta\right) + iy\left(\sin2\varphi +\sin2\theta\right)\right].
    \label{B.2}
\end{align}
where $m$ is an integer number. The next step is to introduce the change of variables $\varphi + \theta = 2\psi$, $\varphi - \theta = 2\chi$ with Jacobian equals to 2. Whereas, the integration limits for $\psi$ and $\chi$ can be chosen as $\left[-\pi,\pi\right]$ and $\left[0,\pi\right]$, correspondingly, due to the symmetry of the integrand \cite{Watson1966}. Thus, we have
\begin{align}
\mathcal{J}_{n-m}\left(x,y\right)\mathcal{J}_{n+m}\left(x,y\right) &= \nonumber \\
&\hspace{-3cm} \frac{1}{2\pi^2}\int\limits_0^\pi d\chi\exp\left(2im\chi\right) \int\limits_{-\pi}^{\pi}d\psi\exp\left(2ix\cos\chi\sin\psi +2iy\cos2\chi\sin2\psi -2in\psi\right).
    \label{B.3}
\end{align}
We recognize in the inner integral on the right-hand side of (\ref{B.3}) the integral representation of the generalized Bessel function (\ref{B.1}) with the new arguments and the doubled index. Hence, we can write
\begin{align}
\mathcal{J}_{n-m}\left(x,y\right)\mathcal{J}_{n+m}\left(x,y\right) = \frac{1}{\pi}\int\limits_0^\pi d\chi\cos\left(2m\chi\right)\mathcal{J}_{2n}\left(2x\cos\chi,2y\cos2\chi\right),
    \label{B.4}
\end{align}
where we also exploited the fact that the left-hand side of (\ref{B.3}) is a real-valued function for real $x$ and $y$. For $m=0$ and $m = 1$ we have integrals that appear in the calculations of LP case (see Eqs.~\eqref{Eq.2.19} and \eqref{Eq.2.20}).

\bibliography{references}

\end{document}